\documentstyle[12pt, amsmath, amsfonts, epsfig, amssymb, rotating]{article}
\begin{document}
\def\be{\begin{equation}}
\def\ovx{\vec{x}}
\def\ovxo{\vec{x}_0}
\def\ovy{\vec{y}}
\def\ovr{\vec{r}}
\def\ovn{\vec{n}}
\def\vp{\varphi}
\def\eps{\epsilon}
\def\call{{\mathcal L}}
\def\calv{{\mathcal V}}
\def\calw{{\mathcal W}}
\def\calo{{\mathcal O}}
\def\calp{{\mathcal P}}
\def\calpo{{\mathfrak o}}
\def\callo{{\mathcal L}_{0}}
\def\calg{{\mathcal G}}
\def\calgo{{\mathcal G}_{0}}
\def\rm{{\mathbb R}}
\def\nm{{\mathbb N}}
\def\cm{{\mathbb C}}
\def\zm{{\mathbb Z}}
\def\hm{{\mathbb H}}
\def\re{{\mathcal R}e\ }
\def\im{{\mathcal I}m\ }
\def\equi{\displaystyle{\ \ \mathop{\sim}_{s\to 1}\ \ }}
\def\da{\partial_1}
\def\db{\partial_2}
\def\di{\partial_i}
\def\jt{\tilde{J}}
\def\ensdn{D_{-}}
\def\ensdp{D_{+}}
\def\sgn{\ \textrm{sgn}\ }
\def\tlog{\widetilde{\log}}
\def\vsp{v_{\textrm{sp}}}
\def\usp{u_{\textrm{sp}}}
\def\ba{\bar{\alpha}}
\def\an{\frac{\alpha}{N}}
\def\ei{\textrm{Ei}}
\def\eun{E_{1}}
\def\binom{\displaystyle{C_{N}^{n}}}
\def\vpint{{\mathbf -}\hspace{-4.4mm}\int}

\title{\bf Nearest-neighbor distribution for singular billiards}
\author{E. Bogomolny, O. Giraud, and C. Schmit\\
 Laboratoire de Physique Th\'eorique et Mod\`eles Statistiques 
 \thanks{Unit\'e Mixte de Recherche de l'Universit\'e Paris XI et du CNRS
   (UMR 8626)}\\
Universit\'e de Paris XI, B\^at. 100\\
91405 Orsay Cedex, France}

\maketitle

\begin{abstract}

The exact computation of the nearest-neighbor spacing distribution $P(s)$
is performed for a rectangular billiard with point-like scatterer inside
for periodic and Dirichlet boundary conditions and  it is demonstrated that
when $s\rightarrow \infty$ this function decreases exponentially.  
Together with the results of Ref.~\cite{singular} it  proves that spectral 
statistics of such systems is of intermediate type characterized by 
level repulsion at
small distances and exponential fall-off of the nearest-neighbor
distribution at large distances. The calculation of the $n$-th
nearest-neighbor spacing distribution $P_{n}(s)$ and its asymptotics 
is performed as well for any boundary conditions. 

\end{abstract}

\pagebreak

\section{Introduction}

The statistical analysis of quantum energy levels for a given system in the
semi-classical limit is a well-studied feature in the theory of spectral 
statistics \cite{porter}-\cite{bohigas}. The main conjectures in this field
are the follow: 

(i) The Berry-Tabor conjecture \cite{berry349}:
generic integrable systems obey Poisson statistics, which implies that their 
energy levels behave as independent random variables. 

(ii) The Bohigas-Giannoni-Schmit conjecture \cite{bgc}: generic chaotic systems
follow the Wigner-Dyson distributions of Random Matrix Theory (see \cite{mehta}).

There is enormous amount of numerical evidences that many physical systems do 
agree with these two main  level statistics. 
Partial analytical results support these conjectures for integrable rectangular 
billiards \cite{marklof} and quantum chaotic systems 
(\cite{andreev1}-\cite{keating}).

However, there exist systems which are neither integrable nor chaotic and 
their spectral statistics do not follow any of the above  leading models. In
many cases  their statistics  have  features intermediate between the Poisson
statistics and that of random matrix ensembles and for this reason they are called
``intermediate statistics'' \cite{shklovskii}, \cite{intermediate},
\cite{singular}. 
For the first time this type of behavior had clearly been observed numerically
for the 3-dimensional Anderson model at the metal-isolator transition point
\cite{shklovskii} and later it was argued \cite{ugerland} that spectral statistics
of diffractive and pseudo-integrable systems is also of intermediate type.

To study precisely the statistical behavior of the energy levels of
quantum systems one usually introduces different functions that characterize
the statistics \cite{mehta}. The most important quantity  for our purpose is
the distribution of nearest-neighbor spacings, $P(s)$, which is the
probability that two levels are separated by a distance $s$ with no
levels inside this interval.  

For the Poisson statistics the
nearest-neighbor distribution takes the following particularly simple form 
\be
P(s)=\exp(-s)
\end{equation}
and it is characterized by (i) the absence of  level repulsion ($P(0)\neq 0$)
and (ii) exponential decay for large distance.

For standard random matrix ensembles 
the nearest-neighbor spacing distributions are given by complicated
expressions  \cite{mehta} but their main features can be seen from the Wigner
surmise
\be
P(s)=a_{\beta} s^{\beta} \exp(-c_{\beta} s^{2}),
\end{equation}
where $\beta=1, 2$ and $4$ corresponds, respectively, to  orthogonal, unitary or
symplectic ensembles, and $a_{\beta}$ and $c_{\beta}$ are constants determined
by the normalization conditions. Its main properties are (i)  level
repulsion, $P(0)=0$, and (ii) very quick decrease at large values of $s$,
$P(s)\propto \exp ( -cs^2)$ when $s\rightarrow \infty$.

We call spectral statistics of intermediate type if they have the following
hybrid properties (cf. \cite{shklovskii}, \cite{ugerland}): 
(i) they exhibit the level repulsion, $P(0)=0$, as for standard random matrix
ensembles, and (ii) they have exponential decay at large $s$,
$P(s)\propto \exp (-c s)$ when $s\rightarrow \infty$, similarly to the
Poisson statistics.
Little is known analytically for systems with intermediate statistics
though it is possible to write down models which will have this type of
statistics \cite{gaudin}, \cite{intermediate}.  

The rectangular billiard with a point-like scatterer inside belongs to the class 
of diffractive systems and is one of the models which is supposed to have 
intermediate statistics \cite{ugerland}, \cite{singular}.
Without the scatterer this model is an integrable system and when the ratio 
$a^2/b^2$ of the sides of the rectangle is a `good' irrational number
its quantum energy  levels $\{e_n\}$  obey the  Poisson statistics 
\cite{berry349},  \cite{marklof}. The addition of a $\delta-$function
scatterer 
\be
\label{potentiel}
V=\lambda\ \delta(\ovx-\ovxo)
\end{equation} 
inside the  rectangle corresponds to a rank-one perturbation, and  the new
quantum energy levels $E$ of the perturbed rectangular billiard will obey 
the following quantization condition \cite{albeverio}, \cite{seba}
\be
\label{equationvp}
\lambda \sum_{n} \frac{|\psi^{(0)}_{n}(\ovxo)|^{2}}{E-e_n}=1,
\end{equation}
where $\psi^{(0)}_{n}$ and $e_n$ are the eigenfunctions and eigenvalues of
the unperturbed rectangular billiard. Similar equations appear in different
models. In particular, the quantization condition for the star graphs can be
transformed to this form \cite{Berko}. All our results are applicable without
changes in such cases as well.

Rank-one perturbations have been
studied in the context of ballistic motion of particles in regular  
\cite{seba} or chaotic cavities \cite{sieber}, and in the context of random
matrix theory  \cite{aleiner}. When a  $\delta-$function potential
(\ref{potentiel}) is added to a chaotic system  with random matrix statistics 
it has been proved \cite{bolesc} that the new eigenvalues in 
Eq.~(\ref{equationvp}) are also 
distributed  according to the same statistics. In the chaotic case the 
universal part of the spectral statistics is not changed by finite-rank  perturbation.
On the contrary, when the unperturbed system is integrable, the perturbation 
(\ref{potentiel}) changes dramatically its spectral statistics \cite{seba},
\cite{singular}.

In Ref.~\cite{singular} the two-point correlation function of a rectangular billiard 
with a small-size scatterer inside (described by the potential 
(\ref{potentiel})) has been computed analytically. One of the conclusions 
of this paper was that spectral statistics of such singular billiards do have 
level repulsion. For billiards with periodic boundary conditions the
two-point correlation function and, consequently, the nearest-neighbor
distribution vanish linearly at small distances with the slope independent
on the coupling constant
\be
P(s)\stackrel{s\rightarrow 0}{\rightarrow} \frac{\pi\sqrt{3}}{2}s.
\end{equation}
For billiards with Dirichlet boundary conditions the small-$s$ behavior of 
the two-point correlation function and the nearest-neighbor distribution 
is different: one has (see \cite{singular})
\be
P(s)\stackrel{s\rightarrow 0}{\rightarrow} \frac{1}{8\pi^3}s\log^4 s.
\end{equation}
The main purpose of this paper is to compute analytically the 
nearest-neighbor spacing distributions for this model and demonstrate
that any boundary conditions they decrease exponentially at large
separation. Together with the results of Ref.~\cite{singular} about 
the existence of  level repulsion it will furnish the proof that 
spectral statistics of these systems are of intermediate type.

The plan of the paper is following. In Section \ref{genefor} we generalize
the formalism used in \cite{singular} to describe the nearest-neighbor spacing 
distribution for a billiard with a point-like scatterer for periodic and
Dirichlet boundary conditions. Though the resulting formulas are explicit
and exact they are quite cumbersome and in Section~\ref{asymbeh}  we study the
asymptotic behavior of $P(s)$ for large $s$. It is demonstrated that in all
cases the nearest-neighbor distribution has exponential tail at large
distances  thus proving the intermediate character of spectral statistics
of singular billiards. In Section~\ref{nnearest} the $n$-th nearest-neighbor
spacing distributions for these billiards with  periodic and  Dirichlet
boundary conditions are computed analytically and their large distance
asymptotics are found as well. In Appendix we present certain technical 
details of the computation of necessary integrals.

\section{The general formalism}\label{genefor}

\subsection{Preliminary computations}
In this section, our aim is to find analytical expressions for the
nearest-neighbor spacing distribution of the solutions, $E$, of the
following  equation
\be
\label{equN}
\lambda \sum_{j=1}^{N} \frac{r_j}{E-e_j}=1
\end{equation}
where $e_j$ $j=1,\ldots,N$  are independent
random variables with a uniform distribution $d\mu(e)$:
\be
d\mu(e)=\left \{ \begin{array}{ll}\frac{1}{2W}de&   \textrm{if} -W\leq e\leq W  \cr
0&    \textrm{otherwise}  \end{array} \right. ,
\end{equation}  
and $r_j$ are positive constants with  mean value 1
\be
\frac{1}{N}\sum_{n=1}^Nr_n=1.
\end{equation}
This normalization condition permits to introduce conveniently the coupling
constant $\lambda$.

In general this equation describes zeros of  a meromorphic function whose
poles are assumed to be independent random variables and it can correspond
to different physical problems (see e.g. \cite{delande}).
In this paper we prefer to consider it as the quantization condition of
rectangular (or more general integrable) billiards with a small-size
impurity inside  \cite{albeverio}, \cite{seba}. To ensure that energy levels
of unperturbed billiards behave as independent random variables it is
necessary to assume that the ratio of squares of the sides of the 
rectangle, $a$
and  $b$,  is an irrational number badly approximated by a rational 
(that is a diophantine number) \cite {marklof} with the following property
\begin{equation}  
|\frac{a^2}{b^2}-\frac{m}{n}|>\frac{C}{n^k}
\end{equation}
for all integers $m$, $n$ and some $k\geq 2$.

The residues, $r_n$, depend on boundary conditions. For quantum problems with
periodic boundary conditions $r_n=1$. For Dirichlet conditions
\begin{equation}
r_{mn}=4\sin^2 \frac{\pi}{a}mx_0 \sin^2 \frac{\pi}{b}my_0
\label{res}
\end{equation}
where $x_0$, $y_0$ are coordinates of the singular scatterer. When the ratios
$x_0/a$ and  $y_0/b$ are non-commensurable irrational numbers and
$m,n\rightarrow \infty$ $r_{mn}$ can be considered as independent random
variables
\begin{equation}
r_{mn}=4 \sin^2 \phi_1 \sin^2 \phi_2
\label{rdir}
\end{equation}
with angles $\phi_i$ uniformly distributed between 0 and $\pi/2$. 

When both ratios $x_0/a$ and  $y_0/b$ are rational numbers
\begin{equation}
\frac{x_0}{a}=\frac{p_1}{q_1},\;\; \frac{y_0}{b}=\frac{p_2}{q_2}
\label{pq}
\end{equation}
with co-prime integers $(p_i,q_i)$ the residues (\ref{res}) only depend  on
$m$ mod $q_1$ and $n$ mod $q_2$ and there is only finite number of 
residues determined by $q_i$ angles $\phi_i$ in (\ref{rdir}) 
(see \cite{singular} for more detail)
\begin{eqnarray}
\phi_1&=&\pi\frac{k_1}{q_1},\;\mbox{with}\;k_1=0,1,\ldots, q_1-1;\nonumber\\ 
\phi_2&=&\pi\frac{k_2}{q_2},\;\mbox{with}\;k_2=0,1,\ldots, q_2-1. 
\label{rational}
\end{eqnarray}
All our formulas below remain valid for general $r_n$.

Obviously there are
$N$ solutions $E_j$ of Eq.~$(\ref{equN})$ since each interval
$]e_i, e_{i+1}[$ contains one and only one of these
solutions. We are interested in the
nearest-neighbor distribution, $P(s)$, that is the probability  that two
energy levels $E_i$ and $E_j$ are neighbors separated by a distance $s$.
In our case it is the probability that 2 solutions  $E_i$ and
$E_j$ of $(\ref{equN})$ are separated by one and only one unperturbed level
$e_k$, and that
$|E_i-E_j|=s$. Let us compute at first the probability $P(E_1, E_2)$ that
two given energy levels $E_1$ and $E_2$ be neighbors. Assuming for instance
that $E_2<E_1$,  $P(E_1, E_2)$ is the probability that that one solution of
$(\ref{equN})$ equals $E_1$, another one equals $E_2$, and that there exists
$i$, $1\leq i \leq N$, such that 
\begin{eqnarray}
\left\{\begin{array}{l}
  e_i\in   \left]E_2, E_1\right[\cr
\forall j\neq i, e_j \notin  \left]E_2, E_1\right[.
\end{array}\right.
\end{eqnarray}
As it is supposed that $e_k$ are independent
random variables with a uniform distribution, 
\be
\label{probe1e2}
P(E_1, E_2)=\int_{-W}^{W}\prod_{k=1}^N \frac{de_k}{2W}
\rho(E_1)\rho(E_2)\sum_{i=1}^{N}
\chi(e_i)\prod_{j\neq i}\left(1-\chi(e_j)\right),
\end{equation}
where $\chi(e)$ is the characteristic function of the interval $]E_2, E_1[$ equal 
to $1$ if $e$ belongs to  $]E_2, E_1[$, and to $0$ otherwise, and $\rho$ is the 
density of the solutions $E_i$:
\be
\label{densite}
\rho(E)=\sum_{i=1}^{N}\delta(E-E_i).
\end{equation}
It is convenient to rewrite these formulas in a more symmetric way:
\be
P(E_1, E_2)= \sum_{\{\sigma_k\}}\int\prod_{k=1}^N d\mu_{\sigma_k}(e_k)
\rho(E_1)\rho(E_2),
\label{psigma}
\end{equation}
where variables $\sigma_k$, $k=1,\ldots,N$ take 2 values: 0, 1; and we 
introduce two different measures
\be
\int d\mu_0(e)\phi (e)=\frac{1}{2W}\int_{-W}^{W}\chi(e)\phi (e)de=
\frac{1}{2W}\int_{E_2}^{E_1}\phi (e)de
\label{m0}
\end{equation}
and
\be
\int d\mu_1(e)\phi (e)=\frac{1}{2W}\int_{-W}^{W}(1-\chi(e))\phi (e)de=
\frac{1}{2W}(\int_{-W}^{W}-\int_{E_2}^{E_1})\phi (e)de.
\label{m1}
\end{equation}
The summation in (\ref{psigma}) is performed over all sequences $\sigma_k$
which contain 1 zero and $N-1$ ones.

Because $\{E_{i}\}$ are solutions of Equation $(\ref{equN})$ the density of
states $(\ref{densite})$ can be rewritten under the form (cf. \cite{singular})
\be
\label{rhodelta}
\rho(E)=\delta\left(\sum_{i=1}^{N}\frac{r_j}{E-e_j}-\frac{1}{\lambda}\right)
\sum_{k=1}^{N}\frac{r_k}{(E-e_k)^{2}}.
\end{equation}
Representing the $\delta-$function by a Fourier integral one gets
\be
\rho(E)=\int_{-\infty}^{\infty}\frac{d\alpha}{2\pi}
\exp\left(i\alpha\left(\sum_{i=1}^{N}\frac{r_j}{E-e_j}-\frac{1}{\lambda}\right)\right)
\sum_{k=1}^{N}\frac{r_k}{(E-e_k)^{2}}
\end{equation}
and finally the probability $(\ref{probe1e2})$ can be put under the form
\begin{eqnarray}
\label{premierps}
&P(E_1, E_2)=\displaystyle{\int_{-\infty}^{\infty}\frac{d\alpha_1 d\alpha_2}{4\pi^2}
\sum_{\{\sigma_k\}}\int\prod_{k=1}^Nd \mu_{\sigma_k}(e_k)
e^{-i(\alpha_1 +\alpha_2)/\lambda}}\\
&\times \displaystyle{\sum_{k_1,k_2=1}^N 
\frac{r_{k_1} r_{k_2}}{(E_1-e_{k_1})^2 (E_2-e_{k_2})^2}
\prod_{j_1, j_2=1}^{N}
 \exp\left(i \alpha_1\frac{r_{j_1}}{E_1-e_{j_1}}+i
  \alpha_2\frac{r_{j_2}}{E_2-e_{j_2}}\right)}.\nonumber
\end{eqnarray}
Let us introduce the following functions:
\begin{eqnarray}  
f_{\sigma}(\alpha_1,\alpha_2)&=&\int d\mu_{\sigma}(e)
  \exp\left(i\frac{\alpha_1}{E_1-e}+i\frac{\alpha_2}{E_2-e}\right),\nonumber\\
\Psi_{j\sigma}(\alpha_1,\alpha_2)&=&\int d\mu_{\sigma}(e)
  \frac{1}{(E_j-e)^2}
\exp\left(i\frac{\alpha_1}{E_1-e}+i\frac{\alpha_2}{E_2-e}\right),
\label{deffunctions}\\
g_{\sigma}(\alpha_1,\alpha_2)&=&\int d\mu_{\sigma}(e)
  \frac{1}{(E_1-e)^2 (E_2-e)^2}
\exp\left(i\frac{\alpha_1}{E_1-e}+i\frac{\alpha_2}{E_2-e}\right).\nonumber
\end{eqnarray}
The nearest-neighbor distribution can be expressed through these functions 
in the following way
\begin{eqnarray}
P(E_1, E_2)&=&\sum_{\{\sigma_k\}}
\displaystyle{\int_{-\infty}^{\infty}\frac{d\alpha_1 d\alpha_2}{4\pi^2}}
\left (\sum_j r_j^2 g_{\sigma_j}(\alpha r_j)
\prod_{k\neq j} f_{\sigma_k}(\alpha r_k)\right . \label{ABCD}\\
&+&\left .\sum_{j\neq k} r_j r_k \Psi_{1\sigma_j}(\alpha r_j)
\Psi_{2\sigma_k}(\alpha r_k)
\prod_{l\neq j,k}f_{\sigma_l}(\alpha r_l) \right )
e^{-i(\alpha_1  +\alpha_2)/\lambda}.
\nonumber
\end{eqnarray}
Here and below when it will not lead to a confusion we use the notation
$f(\alpha)$ for a function of two variables $f(\alpha_1,\alpha_2)$
and $f(\alpha r)$ instead of $f(\alpha_1 r, \alpha_2 r)$.
This formula is valid for all sequences of $\sigma_k$.
The functions (\ref{deffunctions}) with index $\sigma=0$ correspond
to unperturbed level between $E_2$ and $E_1$, the functions with  index
$\sigma=1$ correspond to unperturbed level outside $] E_2,E_1 [$. Therefore
to describe the nearest-neighbor distribution the summation should be 
done over $N$ possible sequences containing only  1 zero.

The functions $\Psi_{j\sigma}$ and $g_{\sigma}$ are related 
to $f_{\sigma}$ by the relations
\begin{equation}
\label {relpsifg}
\Psi_{j\sigma}(\alpha_1,\alpha_2)=
-\frac{\partial^2}{\partial\alpha_{j}^{2}}f_{\sigma}(\alpha_1,\alpha_2),\;
g_{\sigma}(\alpha_1,\alpha_2)=
\frac{\partial^4}{\partial\alpha_{1}^{2}\partial\alpha_{2}^{2}}
f_{\sigma}(\alpha_1,\alpha_2).
\end{equation}
Therefore in order to compute $P(E_1, E_2)$ it is necessary to find
only $f_{\sigma}$. Let us  introduce the functions
\be
\label{defI}
I_{\sigma}(\alpha_1,\alpha_2)=2W\int d\mu_{\sigma}(e)
[1-\exp (i\frac{\alpha_1}{E_1-e}+i\frac{\alpha_2}{E_2-e})],
\end{equation}
which are related to our basic functions $f_{\sigma}$
as follows:
\begin{eqnarray}
f_1(\alpha_1,\alpha_2)&=&
1-\frac{\omega}{2W}-\frac{1}{2W}I_1(\alpha_1,\alpha_2), 
\nonumber\\
f_0(\alpha_1,\alpha_2)&=&\frac{\omega}{2W}-\frac{1}{2W}I_0(\alpha_1,\alpha_2),
\label{f1}
\end{eqnarray}
where $\omega=E_1-E_2$ is the difference of energies (we recall that we
have assumed $E_2<E_1$).

The integral defining $I_1(\alpha)$ can be split into two parts
\begin{eqnarray}
I_1(\alpha_1,\alpha_2)&=&(\int_{-W}^{E_2}+\int_{E_1}^{W})
[1-\exp (i\frac{\alpha_1}{E_1-e}+i\frac{\alpha_2}{E_2-e})]de\nonumber\\
&=&J_1(\alpha_1,\alpha_2)+j(\alpha_1,\alpha_2),
\label{split}
\end{eqnarray}
where
\be
J_1(\alpha_1,\alpha_2)=(\int_{-\infty}^{E_2}+\int_{E_1}^{\infty})
[1-\exp (i\frac{\alpha_1}{E_1-e}+i\frac{\alpha_2}{E_2-e})]de 
\label{J1}
\end{equation}
and
\be
j(\alpha_1,\alpha_2)=-(\int_{-\infty}^{-W}+\int_{W}^{\infty})
[1-\exp (i\frac{\alpha_1}{E_1-e}+i\frac{\alpha_2}{E_2-e})]de.
\label{jint}
\end{equation}
For convenience we define the function $J_0(\alpha)=I_0(\alpha)$ so that
from (\ref{defI}) 
\be
J_0(\alpha_1,\alpha_2)=\int_{E_2}^{E_1}
[1-\exp (i\frac{\alpha_1}{E_1-e}+i\frac{\alpha_2}{E_2-e})] de.
\label{J0}
\end{equation}
The integral (\ref{jint}) defining $j(\alpha)$ has no singularity inside 
the integration region and as it was demonstrated in \cite{singular} it is
sufficient to take into account only terms linear in $\alpha$  and to ignore the
difference between $E_1$ and $E_2$ (i.e. set $E_1\approx E_2\approx E$). 
In this approximation 
\be
j(\alpha_1,\alpha_2)= i(\alpha_1+\alpha_2)\log \frac{W-E}{W+E}.
\end{equation}
On the contrary the functions $J_{\sigma}(\alpha)$ are quite cumbersome.
One can easily check that they depend only on the difference of energies,
$\omega=E_1-E_2$,  and  that 
\be
J_{\sigma}(\alpha)=\omega \tilde{J}_{\sigma}(\frac{\alpha}{\omega})
\label{scaling}
\end{equation}
where the functions $\tilde{J}_{\sigma}(\alpha)$ are defined by
Eqs.~(\ref{J1}) and (\ref{J0}) with $E_1=1$ and $E_2=0$.

In Appendix it is demonstrated that these functions obey the differential
equation
\be
\label{dadbj}
(\da-\db)\tilde{J}_{\sigma}(\alpha_1,\alpha_2)=
e^{i(\alpha_1-\alpha_2)}\phi_{\sigma}(\alpha_1,\alpha_2)
\end{equation}
where $\partial_i$ denotes the derivative with respect to $\alpha_i$ and
functions $\phi_{\sigma}(\alpha)$ at real $\alpha$ are given by
(\ref{realphi1}) and (\ref{realphi0}). From this equation it follows (see
Appendix  for details) that the function $\tilde{J}_1(\alpha)$ is an
analytical function of 2 complex variables $\alpha_1$, $\alpha_2$ with the
cuts as in Fig.~\ref{cut}a and \ref{cut}b given by the following expression 
\be
\label{finalJ1}
\tilde{J}_1(\alpha_1,\alpha_2)=
\vpint_{-1}^{\infty}\frac{d t}{t^2}
\left(1-e^{i(\alpha_1+\alpha_2) t}\right)
+\int_{0}^{\alpha_1}e^{i(2 t-\alpha_1-\alpha_2)}
\phi_1(t,\alpha_1+\alpha_2-t) d t,
\end{equation}
where $\vpint$ denotes the principal part of the integral. 

The function $\tilde{J}_0(\alpha)$ is an analytical function in a 
region indicated in Fig.~\ref{cut}c and \ref{cut}d with integral
representation
\be
\label{finalJ0}
\tilde{J}_{0}(\alpha_1,\alpha_2)=
\int_{1}^{\infty}\frac{d t}{t^2}
\left(1-e^{-i(\alpha_1+\alpha_2)  t}\right)
+\int_{0}^{\alpha_1}e^{i(2 t-\alpha_1-\alpha_2)}
\phi_{0}(t,\alpha_1+\alpha_2-t) d t.
\end{equation}
\begin{figure}[ht]
\begin{center}
\epsfig{file=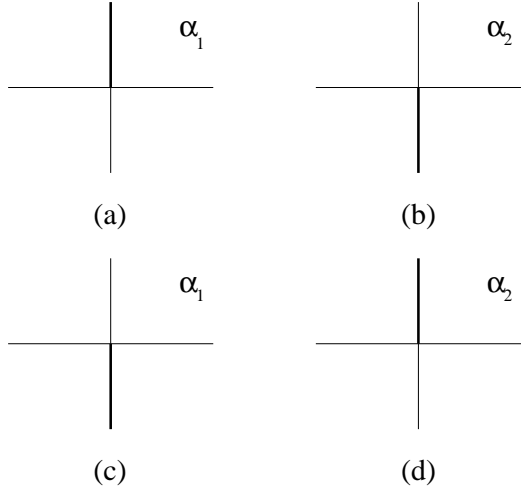, angle=270 ,width=10cm}
\end{center}
\caption{The cuts in complex planes of $\alpha_1$ and $\alpha_2$ for functions
  $\tilde{J}_1$ and $\phi_1$: (a), (b);   $\tilde{J}_0$ and $\phi_0$: 
  (c), (d).}
\label{cut}
\end{figure}

Exactly as it was done in \cite{singular} one can prove that  functions 
$\Psi_{i\sigma}$, 
$g_{\sigma}$ defined in Eqs.~(\ref{deffunctions}) can be expressed 
through the  functions $\phi_{\sigma}$ as follows: 
\begin{eqnarray}
g_{\sigma}(\alpha_1,\alpha_2)&=&
\frac{1}{2W\omega^2}(\da-\db) [e^{i\frac{\alpha_1-\alpha_2}{\omega}}
\phi_{\sigma}(\frac{\alpha_1}{\omega},\frac{\alpha_2}{\omega})],\nonumber\\
\Psi_{1\sigma}(\alpha_1,\alpha_2)&=&
\frac{1}{2W}e^{i\frac{\alpha_1-\alpha_2}{\omega}}\da
\phi_{\sigma}(\frac{\alpha_1}{\omega},\frac{\alpha_2}{\omega}),
\label{exprgpsi}\\
\Psi_{2\sigma}(\alpha_1,\alpha_2)&=&
-\frac{1}{2W}e^{i\frac{\alpha_1-\alpha_2}{\omega}}\db
\phi_{\sigma}(\frac{\alpha_1}{\omega},\frac{\alpha_2}{\omega}).\nonumber
\end{eqnarray}

\subsection{Nearest-neighbor spacing distribution}\label{nnd}

Using Eqs.~$(\ref{exprgpsi})$ one can integrate the first term in (\ref{ABCD})
by parts and because (see (\ref{f1}) and (\ref{scaling}))
\be
(\da-\db)f_{\sigma}(\alpha_1,\alpha_2)=-\frac{1}{2W}
e^{i(\alpha_1-\alpha_2)/\omega}
\phi_{\sigma}(\frac{\alpha_1}{\omega},\frac{\alpha_2}{\omega})
\end{equation}
one obtains
\begin{eqnarray}
& &\displaystyle{\int_{-\infty}^{\infty}\frac{d\alpha_1 d\alpha_2}{4\pi^2}}
\sum_j r_j^2 g_{\sigma_j}(\alpha r_j)
[\prod_{k\neq j} f_{\sigma_k}(\alpha r_k)]e^{-i(\alpha_1+\alpha_2)/\lambda}
\nonumber\\
&=&\frac{1}{\omega^2(2W)^2}
\displaystyle{\int_{-\infty}^{\infty}\frac{d\alpha_1 d\alpha_2}{4\pi^2}
\sum_{j\neq k}r_jr_k
\phi_{\sigma_j}(\frac{\alpha}{\omega}r_j)
\phi_{\sigma_k}(\frac{\alpha}{\omega}r_k)
e^{i\frac{\alpha_1-\alpha_2}{\omega}(r_k+r_j)}}
\nonumber\\
&\times&[\prod_{l\neq j,k} f_{\sigma_l}(\alpha r_l)]
e^{-i(\alpha_1+\alpha_2)/\lambda}.
\end{eqnarray}
According to Eqs.~(\ref{exprgpsi})  the second term term in (\ref{ABCD}) can
also be expressed through the same functions $\phi_{\sigma}(\alpha)$ and
after the scaling of variables $\alpha\rightarrow \alpha \omega$ (cf.
(\ref{scaling})) the nearest-neighbor distribution (\ref{ABCD}) takes the form
\begin{eqnarray}
P(\omega)&=&
\sum_{\{\sigma_{k}\}}
\displaystyle{\int_{-\infty}^{\infty}\frac{d\alpha_1 d\alpha_2}{(4W\pi)^2}}
\sum_{j\neq k}r_jr_k\{ \phi_{\sigma_j}(\alpha r_j)\phi_{\sigma_k}(\alpha r_k)
\nonumber\\  
&-&\da \phi_{\sigma_j}(\alpha r_j)\db \phi_{\sigma_k}(\alpha r_k)\} 
e^{i(\alpha_1-\alpha_2)(r_j+r_k)}
\nonumber\\
&\times&[\prod_{l\neq j,k} \tilde{f}_{\sigma_l}(\alpha r_l)]
e^{-i\omega (\alpha_1+\alpha_2)/\lambda},
\label{pomega}
\end{eqnarray}
where $\tilde{f}_{\sigma_l}(\alpha )=f_{\sigma_l}(\alpha \omega)$. Using
Eqs.~(\ref{f1}), (\ref{split}), and (\ref{jint}) one gets
\begin{eqnarray}
\tilde{f}_1(\alpha_1,\alpha_2)&=&
1-\frac{\omega}{2W}(1+\tilde{J}_1(\alpha_1,\alpha_2)+i(\alpha_1+\alpha_2)
\log\frac{W-E}{W+E}) ,
\nonumber\\
\tilde{f}_0(\alpha_1,\alpha_2)&=&\frac{\omega}{2W}(1-\tilde{J}_0(\alpha_1,\alpha_2)).
\label{barf}
\end{eqnarray}
Expression (\ref{pomega}) is valid for any sequence 
$\{\sigma_k\}\in\{0,1\}^N$. To get the
nearest-neighbor distribution one has to sum over $N$ sequences containing
only one zero. Taking into account that in the limit $N\rightarrow \infty$
the restriction $j\neq k$ is unessential we obtain, keeping only the
dominant term,
\begin{eqnarray}
P(\omega)&=&
\displaystyle{\frac{N^2}{4W^2}\int_{-\infty}^{\infty}
  \frac{d\alpha_1 d\alpha_2}{(2\pi)^2}}
[N<\tilde{f}_0(\alpha r)>V_{\phi_1,\phi_1}(\alpha)+V_{\phi_0,\phi_1}(\alpha)
+V_{\phi_1,\phi_0}(\alpha)]
\nonumber\\
&\times& [\prod_{l} \tilde{f}_{1}(\alpha r_l)]
e^{-i\omega (\alpha_1+\alpha_2)/\lambda},
\end{eqnarray}
where the  operator $V_{f,g}(\alpha)$ is defined for arbitrary functions $f(\alpha)$ 
and $g(\alpha)$ by the following expression
\begin{eqnarray}
V_{f,g}(\alpha)&=& <rf(\alpha r)e^{i(\alpha_1-\alpha_2)r}>
<rg(\alpha r)e^{i(\alpha_1-\alpha_2)r}>
\nonumber\\
&-&<(\frac{\partial}{\partial \alpha_1}f(\alpha r))e^{i(\alpha_1-\alpha_2)r}>
<(\frac{\partial}{\partial \alpha_2}g(\alpha r))e^{i(\alpha_1-\alpha_2)r}>,
\label{vfg}
\end{eqnarray}
and $<f(r)>$ means the mean value over all values of $r$
\be
<f(r)>=\frac{1}{N}\sum_{n=1}^N f(r_n).
\end{equation}
Measuring the energy difference $\omega$ in the units of mean level
spacing 
\be
s=\frac{N\omega}{2W}
\end{equation}
the product $\prod_{l} \tilde{f}_{1}(\alpha r_l)$ can also be simplified in
the limit of large $N$ (see (\ref{barf}))
\be
\prod_{l=1}^N \tilde{f}_{1}(\alpha r_l)\approx 
\exp (-\frac{N\omega}{2W}[1+\tilde{J}_1(\alpha_1,\alpha_2)
+i(\alpha_1+\alpha_2)\log\frac{W-E}{W+E}]).
\end{equation}
Introducing the renormalized coupling constant $\lambda'$
\be
\label{lprime}
\frac{1}{\lambda'}=\frac{2W}{N \lambda}+\log\frac{W-E}{W+E}
\end{equation}
we obtain the final formula for the nearest-neighbor distribution
$P(s)=(2W/N)^2P(\omega)$:
\begin{eqnarray}
&&P(s)=e^{-s}\int_{-\infty}^{\infty}\frac{d\alpha_1 d\alpha_2}{(2\pi)^2}
[s(1-<\tilde{J}_0(\alpha r)>)V_{\phi_1,\phi_1}(\alpha)
\nonumber\\
&&+V_{\phi_0,\phi_1}(\alpha)+V_{\phi_1,\phi_0}(\alpha)]
e^{-s(<\tilde{J}_1(\alpha r)>+i(\alpha_1+\alpha_2)/\lambda')}.
\label{psfinal}
\end{eqnarray}

\subsection{Analytical continuation}\label{analytical}

Usually,  if one wants to compute an integral
\be
\int_{-\infty}^{\infty}d\alpha_1d\alpha_2 f(\alpha_1,\alpha_2)
e^{-sJ(\alpha_1,\alpha_2)},
\label{ints}
\end{equation}
where $J(\alpha)$ and $f(\alpha)$ are analytical functions in a certain
region, the first step is to move the integration contour as far as possible to
decrease the integrand. In general, during that deformation one can either
meet a saddle point or a singularity which signifies that further
deformation of the contour either will increase the integrand or is not
possible. If no  such obstacle appears the integral is zero. 

In the case of the nearest-neighbor distribution (\ref{psfinal}) the 
saddle-point equation reads (taking here $r=1$)
\be
\frac{\partial}{\partial
  \alpha_1}\tilde{J}_1(\alpha_1,\alpha_2)
+\frac{i}{\lambda'}=0,\;\;\;
\frac{\partial}{\partial \alpha_2}\tilde{J}_1(\alpha_1,\alpha_2)
+\frac{i}{\lambda'}=0.
\end{equation}
In particular, these equations  imply that at any saddle-point
\be
(\frac{\partial }{\partial \alpha_1} -
\frac{\partial }{\partial \alpha_2})\tilde{J}_1(\alpha_1,\alpha_2)=0.
\end{equation}
From (\ref{difference}) and (\ref{phi1}) it follows that this difference
is proportional to $K_{0}(x)$ with $x=2\sqrt{\alpha_1 \alpha_2}$.
But  $K_{0}(x)$ has
no zero on the complex plane (see \cite{bateman} p. 62). Therefore
our integral (\ref{psfinal}) has no saddle-points and  one can move the
contour of integration freely. If the pre-factor in (\ref{ints})
has no singularities in
the region where $\tilde{J}_1$  is analytical the contribution vanishes. Note that it is
exactly what should be expected for physical reasons: replacing the pre-factor
in (\ref{psfinal}) by $1$  we have  to obtain the probability that there
are  two perturbed levels in $E_2$ and $E_1$ but  no unperturbed energy 
levels between, which according to Eq.~(\ref{equN}) is impossible. 
These considerations prove that  
the integral (\ref{psfinal}) with a pre-factor equal to $1$ or, more generally,
with any pre-factor analytical in the same domain as $\tilde{J}_1(\alpha)$ 
(and not too quickly increasing on infinity) must vanish. 
  
These arguments allow to simplify considerably the expression 
(\ref{psfinal}) for the nearest-neighbor distribution. 
The pre-factor in Eq.~(\ref{psfinal}) is
\be
f(\alpha)=s(1-<\tilde{J}_0(\alpha r)>)V_{\phi_1,\phi_1}(\alpha)
+V_{\phi_0,\phi_1}(\alpha)+V_{\phi_1,\phi_0}(\alpha).
\end{equation}
First, as only the functions with index 0 have analytical properties different
from that of $\tilde{J}_1(\alpha)$ (see Fig.~\ref{cut}), one can drop the first term
and keep only
\be
f(\alpha)=-s<\tilde{J}_0(\alpha r)>V_{\phi_1,\phi_1}(\alpha)
+V_{\phi_0,\phi_1}(\alpha)+V_{\phi_1,\phi_0}(\alpha).
\label{prefactor}
\end{equation}
Secondly, according to Eqs.~(\ref{sumphi}) and (\ref{sumj}) 
\be
\phi_{0}(\alpha)=-\phi_{1}(\alpha)+
\pi(\mbox{sgn}(\alpha_1)-\mbox{sgn}(\alpha_2))J_0(2\sqrt{-\alpha_1\alpha_2}),
\end{equation}
and 
\be
\tilde{J}_0(\alpha)=-\tilde{J}_1(\alpha)+
\mbox{sgn}(\alpha_1)R(\alpha)+
\mbox{sgn}(\alpha_2)R^{\dag}(\alpha),
\end{equation}
where the function $R(\alpha)$ is defined in (\ref{R}) and
$R^{\dag}(\alpha)=R^{*}(\alpha_2,\alpha_1)$.

When these expressions are substituted in (\ref{prefactor}) the terms with
index 1 can be dropped  out because they have the same analytical properties as
$\tilde{J}_1(\alpha)$ and, as it has been discussed above, their integrals 
vanish. Finally the pre-factor takes the form
\begin{eqnarray}
f(\alpha)&=&-s\left [\mbox{sgn}(\alpha_1)<R(\alpha r)>+\mbox{sgn}(\alpha_2)
<R^{\dag}(\alpha r)>\right ]V_{\phi_1,\phi_1}(\alpha)\nonumber\\
&+&\pi(\mbox{sgn}(\alpha_1)-\mbox{sgn}(\alpha_2))
(V_{J_0,\phi_1}(\alpha)+V_{\phi_1,J_0}(\alpha))\\
&+&2\pi i\frac{\delta (\alpha_1)}{\alpha_2+i\epsilon}
(<e^{-i\alpha_2 r}>)^2-2\pi i
\frac{\delta (\alpha_2)}{\alpha_1-i\epsilon}(<e^{i\alpha_1 r}>)^2,
\nonumber
\end{eqnarray}
where $J_0$ is the Bessel function $J_0(2\sqrt{-\alpha_1\alpha_2})$. The
last term in this equation appears from the differentiation of
$(\mbox{sgn}(\alpha_1)-\mbox{sgn}(\alpha_2))$ in (\ref{vfg}). The resulting
$\delta$-function allows us to take the remaining terms at small $\alpha$.
We also write  $\alpha_{1,2}\mp i\epsilon$ where
$\epsilon\rightarrow 0^{+}$ to remind the region where the functions 
are defined.  

When this expression is substituted in Eq.~(\ref{psfinal}) one gets the
final formula for the nearest-neighbor distribution which can be
conveniently written as a sum of 3 terms
\begin{equation}
P(s)=\frac{e^{-s}}{4\pi^2}(A(s)+B(s)+C(s)),
\label{mainps}
\end{equation}
where
\begin{eqnarray}
&A(s)=-s\int_0^{\infty} d\alpha_1 \int_{-\infty}^{\infty}d\alpha_2
<R(\alpha r)>V_{\phi_1,\phi_1}(\alpha)
e^{-s(<\tilde{J}_1(r\alpha)>+i(\alpha_1+\alpha_2)/\lambda')}
\nonumber\\
&-s\int_0^{\infty} d\alpha_2 \int_{-\infty}^{\infty}d\alpha_1
<R^{\dag}(\alpha r)>V_{\phi_1,\phi_1}(\alpha)
e^{-s(<\tilde{J}_1(r \alpha)>+i(\alpha_1+\alpha_2)/\lambda')}+c.c.\;,
\label{A}
\end{eqnarray}
\begin{eqnarray}
&B(s)=\pi\int_0^{\infty} d\alpha_1 \int_{-\infty}^{\infty}d\alpha_2
(V_{\phi_1,J_0}(\alpha)+V_{J_0,\phi_1}(\alpha))
e^{-s(<\tilde{J}_1(r\alpha)>+i(\alpha_1+\alpha_2)/\lambda')}
\nonumber\\
&-\pi\int_0^{\infty} d\alpha_2 \int_{-\infty}^{\infty}d\alpha_1
(V_{\phi_1,J_0}(\alpha)+V_{J_0,\phi_1}(\alpha))
e^{-s(<\tilde{J}_1(r\alpha)>+i(\alpha_1+\alpha_2)/\lambda')}+c.c.\;,
\label{B}
\end{eqnarray}
and
\begin{eqnarray}
C(s)=&&2\pi i\int_{-\infty}^{\infty}\frac{d\alpha_2}{\alpha_2+i\epsilon}
(<e^{-i\alpha_2 r}>)^2
e^{-s(<\tilde{J}_1(0,r\alpha_2)>+i\alpha_2/\lambda')}
\nonumber\\
&&-2\pi i\int_{-\infty}^{\infty}\frac{d\alpha_1}{\alpha_1-i\epsilon}
(<e^{i\alpha_1 r}>)^2
e^{-s(<\tilde{J}_1(r\alpha_1,0)>+i\alpha_1/\lambda')}.
\label{C}
\end{eqnarray}
These expressions look  quite complicated but in the next Section we
show that their asymptotics when $s\rightarrow \infty$ can easily be computed.

\section{Asymptotic behavior}\label{asymbeh}

The formulas (\ref{mainps})-(\ref{C}) have been written in such a way that
when the integration is performed from $-\infty$ to $+\infty$ the contour of
integration as a whole  can be shifted into the complex plane. The direction
of such a deformation is different for the integration over $\alpha_1$ and that
over $\alpha_2$. It can conveniently be fixed by the following change of
variables:  $\alpha_1=-iv$ or $\alpha_2=iv$. In the new variable the allowed
deformation of the contour is in  both cases $\mbox{Re}v>0$.

Let us consider first the simple integral (\ref{C}). 
From Eqs.~(\ref{fonctionJ1}) and (\ref{symmetry})
of Appendix  it follows that  $\tilde{J}_1(0,iv)=\tilde{J}_1(-iv,0)=I(v)$
where
\be
I(v)=\vpint_{-1}^{\infty}\frac{d t}{t^2}
(1-e^{-v t})=e^v-1-v\ei (v),
\label{Iv}
\end{equation}
and
\be
\ei (v)=-\vpint_{-v}^{\infty}\frac{dt}{t}e^{-t}
\label{EI}
\end{equation}
is the standard exponential integral (see e.g. \cite{bateman} p. 143). 

Consequently,  after the above change of variables  the integral (\ref{C}) 
takes the form
\begin{eqnarray}
C(s)=&&-2\pi i\int_{-i\infty}^{+i\infty}\frac{dv}{v}
(<e^{vr}>)^2
e^{-s(<I(vr)>-v/\lambda')}
\nonumber\\
&&-2\pi i\int_{-i\infty}^{+i\infty}\frac{dv}{v}
(<e^{v r}>)^2
e^{-s(<I(vr)>+v/\lambda')}.
\label{Cv}
\end{eqnarray}
Now one can  move the contour to the right till it goes through the
saddle point. For the first integral the position of the saddle point,
$v_{+}$, is  defined by the equation
\be
\frac{d}{dv}<I(v_{+}r)>=\frac{1}{\lambda'}
\end{equation}
and for the second integral the saddle point $v_{-}$ is determined from the
similar equation but with changed sign of the coupling constant $\lambda'$
\be
\frac{d}{dv}<I(vr)>=-\frac{1}{\lambda'}.
\end{equation}
As  $I'(v)=-\ei (v)$ the saddle points $v_{\pm}$ are roots of the equation 
\be
<r \ei (v_{\pm}r)>= \mp\frac{1}{\lambda'}
\label{vpm}
\end{equation}
and it is possible to prove that for any real $\lambda'$ there is one and only one 
solution of this equation.  

Expanding the exponent in (\ref{Cv}) in vicinity of the saddle points and
taking into account that $I''(v)=-e^v/v$ and
$<I(r v_{\pm})>\mp v_{\pm}/\lambda'=<e^{r v_{\pm}}>-1$ one gets that in the limit
$s\rightarrow \infty$ the function (\ref{C}) is the sum of contributions
from 2 saddle points 
\be
C(s)=(2\pi)^{3/2}\sum_{i=\pm}\frac{(<e^{r v_{i}}>)^2}{\sqrt{<re^{rv_i}>sv_{i}}}
e^{-s(<e^{rv_{i}}>-1)}.
\label{asympC}
\end{equation}
The saddle point with the smallest value of $<e^{rv}>$,
which we denote by $v_{sp}$, dominates and it should formally be the only
one to be taken into account. 
For finite $\lambda'$ it corresponds to the solution of Eq.~(\ref{vpm}) with
negative right-hand side
\be
<r\ei (r v_{sp})> =-\frac{1}{|\lambda'|}.
\label{vsp}
\end{equation}
Note the appearance of the absolute value of the renormalized coupling
constant. When $\lambda'\rightarrow \infty$ both saddle points $v_{\pm}$ will
give comparable contributions and both should be included.

The asymptotics of the other terms (\ref{A}) and (\ref{B}) can be computed
by similar considerations. These functions are defined as double integrals,
the first one is taken from 0 to $\infty$ and the second from $-\infty$ to $\infty$.
To compute their asymptotic behavior for large 
$s$ the latter integral should be deformed into the complex plane as it was
done above and in the former integral one has to take into account only the
lowest order terms according to expansion (\ref{smallphi}). 

Let us consider the first term in Eq.~(\ref{A}). We need to know the
limiting behavior of the integrand when $\alpha_1\rightarrow 0$ and
$\alpha_2+\alpha_1=iv$ (we prefer to use this deformation instead of the
usual one, $\alpha_2=iv$, to simplify the formulas below). From
Eqs.~(\ref{fonctionJ1}), (\ref{smallphi}), and (\ref{phismall}) it follows
that for real $\alpha$
\be
\tilde{J}_1(\alpha_1,iv-\alpha_1)\stackrel{\alpha_1 \simeq  0}{\rightarrow}
I(v)+e^v[\frac{\pi}{2}-i(2\gamma+\log v -1 +\log \alpha_1)]\alpha_1,
\end{equation}
with $I(v)$ defined by (\ref{Iv}). From (\ref{R}) in this limit
$R(\alpha)\simeq \pi \alpha_1 e^{v}$. The dominant contribution in
$V_{\phi_1,\phi_1}(\alpha)$ comes from the second term in (\ref{vfg}) and one
gets $V_{\phi_1,\phi_1}(\alpha)\simeq -i<e^{vr}>^2/(\alpha_1v)$. Combining
all terms together and changing the variable 
$\alpha_1\rightarrow \alpha /(s<re^{rv}>)$ we find that at large $s$
\begin{eqnarray}
&&A(s)=2\pi i\int_0^{\infty}d\alpha \int_{-i\infty}^{+i\infty}\frac{dv}{v}
(<e^{rv}>)^2 e^{-s(<I(vr)>-v/\lambda')}
\nonumber\\
&&\times e^{-\pi \alpha/2}
  \sin [\alpha(\log \alpha -\log s +g(v))]\;\;
  +\;\;(\lambda'\rightarrow -\lambda'),
\end{eqnarray}
where  
\be
g(v)=\log v+ 2\gamma -1 +2\frac{<re^{rv} \log r> }{<re^{rv}>}-\log <re^{rv}>.
\label{gv}
\end{equation}
The integral over $v$ is an analog of Eq.~(\ref{Cv}) and can be computed 
exactly as above:
\be
A(s)=-(2\pi)^{3/2}\sum_{i=\pm}\frac{(<e^{rv_{i}}>)^2}{\sqrt{<re^{rv_i}>sv_{i}}}
f(\log s-g(v_{i}))e^{-s(<e^{rv_{i}}>-1)},
\label{asympA}
\end{equation}
where the function $f(y)$ is given by the integral
\be
f(y)=\int_0^{\infty}e^{-\pi \alpha/2}\sin [\alpha(\log \alpha -y)]d\alpha .
\label{f}
\end{equation}
One can check that the contribution (\ref{B}) when $s\rightarrow \infty$ is
smaller by a factor $1/s$  with respect to (\ref{asympC}) and (\ref{asympA})
and can be neglected.

Finally, we obtain that the nearest-neighbor distribution, $P(s)$, in the
limit of large $s$ has the following asymptotics
\be
P(s)=\frac{(<e^{rv_{sp}}>)^2}{\sqrt{2\pi<re^{rv_{sp}}>v_{sp}}}
    \frac{e^{-s<e^{rv_{sp}}>}}{\sqrt{s}}[1-f(\log s-g(v_{sp}))].
\label{asympps}
\end{equation}
The saddle point value, $v_{sp}$, depends on the renormalized coupling
constant by Eq.~(\ref{vsp})
\be
-<r\ei (rv_{sp})> =\frac{1}{|\lambda'|}.
\label{saddlepoint}
\end{equation}
When $\lambda$ is very large the contribution of the second saddle point
with reversed sign of right-hand side of this equation should be added. 

In Fig.~\ref{function} the plot of the function $f(y)$ defined in (\ref{f})
is presented. When $y\rightarrow \infty$ this function
goes to zero as  $-1/y$. Therefore the true asymptotics of $P(s)$ is given by
the first term in (\ref{asympps})
\be 
P(s)=\frac{(<e^{rv_{sp}}>)^2}{\sqrt{2\pi<re^{rv_{sp}}>v_{sp}}}
    \frac{e^{-s<e^{rv_{sp}}>}}{\sqrt{s}}.
\label{trueasympps}
\end{equation}
But because in Eq.~(\ref{asympps}) the argument of the function $f(y)$ 
up to the constant (\ref{gv}) is $\log s$  this
decrease is quite slow and at numerically accessible values of $s$ of the
order of 10 (i.e. $y$ of the order of 3-4) as it is evident from 
Fig.~\ref{function} this function gives a noticeable contribution. 
\begin{figure}[ht]
\begin{center}
\epsfig{file=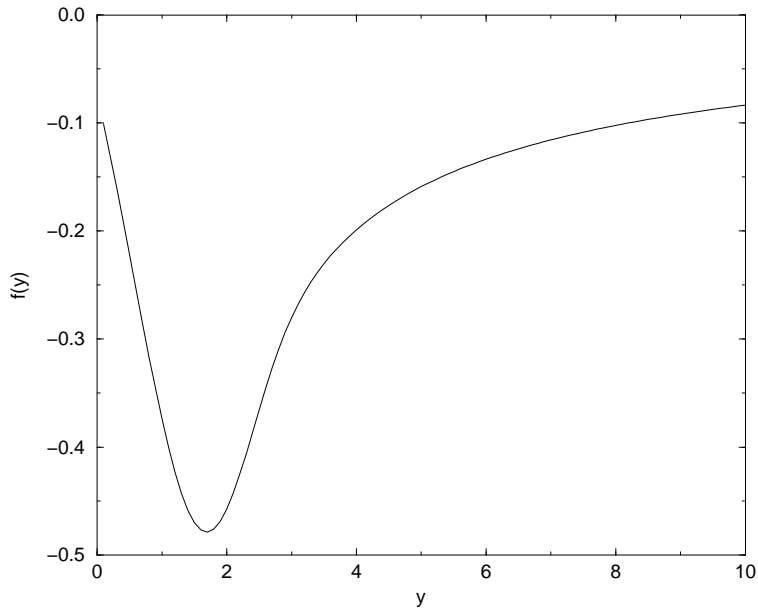 ,angle=-90,width=10cm}
\end{center}
\caption{Plot of the function $f(y)$  defined in (\ref{f}).}
\label{function}
\end{figure}

For the billiard  with periodic boundary conditions  the residues in
(\ref{equN}) all equal 1,  and  all mean values are reduced
to the corresponding function, i.e. for any function $f$ the quantity 
$<f(r x)>$ becomes $f(x)$.

In this case the nearest-neighbor distribution has the following asymptotics
\be
\label{psfini}
P(s)= \sqrt{\frac{e^{3\vsp}}{2\pi\vsp}}\frac{e^{-s e^{\vsp}}}{\sqrt{s}}
[1-f(\log s -g({\vsp}))].
\end{equation}
The value of $\vsp$ is determined by the equation
\be
\ei (\vsp)=-\frac{1}{|\lambda'|},
\label{vspper}
\end{equation}
and 
\be
g(\vsp)=\log \vsp -\vsp +2\gamma -1.
\end{equation}

In Fig.~\ref{n1per2} we present the comparison between numerical
computations and the theoretical prediction $(\ref{psfini})$ for 2 values of
$\lambda$. The logarithm of the nearest-neighbor distribution is plotted as
a function of $s$.  The upper curve (squares) corresponds to $10^{6}$ levels with
$\lambda'=1$ and the lower one (circles) to $5. 10^{5}$ levels with $\lambda'=100$. 
For clarity the curve corresponding to $\lambda'=1$ has been shifted on 
the right by $s\to s+2$. The solid lines represent  theoretical predictions 
(\ref{psfini}) for these values of the coupling constant. For $\lambda'=100$ 
two saddle points with different sign of the right-hand side of
Eq.~(\ref{vspper})
have been taken into account which roughly doubles the result (\ref{psfini}).
\begin{figure}[ht]
\begin{center}
\epsfig{file=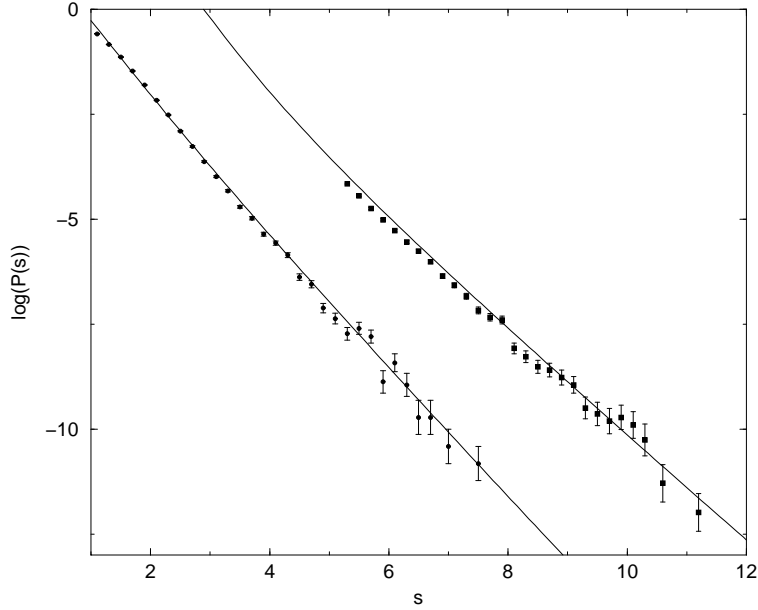 ,angle=-90,width=10cm}
\end{center}
\caption{The nearest-neighbor distribution in the periodic case. Squares
  and circles correspond respectively to $\lambda'=1$  and $\lambda'=100$.
  Solid lines: theoretical predictions (\ref{psfini}). For clarity the upper
  curves are shifted to the right by 2 units.}
\label{n1per2}
\end{figure}
These results  very well confirm theoretical asymptotics of the
nearest-neighbor distribution (\ref{psfini}).

In the case of a rectangular billiard of size $a\times b$ with Dirichlet
boundary conditions, with a point-like scatterer such that the ratios of
its positions $(x_{0}, y_{0})$ to the corresponding sides are
non-commensurable irrational numbers the residues, $r_n$, can be considered
as random variables of the form given by Eq.~(\ref{rdir}) and the mean
value of a given function $f$, $<f(r)>$, should be computed as follows
\be
<f(r)>=\frac{4}{\pi^2}\int_0^{\pi/2}d\phi_1\int_0^{\pi/2}d\phi_2\;
f(4\sin^2\phi_1\sin^2\phi_2).
\label{fmean}
\end{equation}
In the case where $x_0/a$ and $y_0/b$ are rational numbers (\ref{pq}),
the mean value $<f(r)>$ takes the form (see \cite{singular})
\be
<f(r)>=\frac{1}{q_1q_2}\sum_{k_1=0}^{q_1-1}\sum_{k_2=0}^{q_2-1}
f(4\sin^2\frac{\pi k_1}{q_1}\sin^2\frac{\pi k_2}{q_2}).
\label{fmeanrational}
\end{equation}
The cases $k_1=0$ and $k_2=0$ correspond to unperturbed subsequence
of levels (for  Dirichlet boundary conditions) and one has a freedom to
include them to the level density (\ref{densite}) or consider them separatively.
In the later case the terms with $k_i=0$ should be omitted and
$1/(q_1q_2)$ in front of the sum should be substituted by $1/((q_1-1)(q_2-1))$.
The generalizations of $<f(r)>$ for the case when only one ratio  $x_0/a$ or 
$y_0/b$ is an irrational number and the other one is a rational number or both
ratios are irrational but commensurable numbers are straightforward.

With such definition of the mean value the asymptotics of 
the nearest-neigh\-bor distribution is given by Eq.~(\ref{asympps}).

As in the periodic case the quantity $-\log P(s)$ is expected to be nearly
linear with the slope $<e^{r\vsp}>$ where $\vsp$ is the solution of 
Eq.~(\ref{saddlepoint}). 

Fig.~\ref{psndir1} shows the comparison between  numerically computed 
nearest-neighbor distribution for a billiard  with  Dirichlet
boundary conditions  
and the expected asymptotic behavior (\ref{asympps}) (solid line) for
2 values of the renormalized coupling constant, $\lambda'=1$ and 
$\lambda'=10$  ($x_0/a$, $y_0/b$ are incommensurable irrational numbers).
\begin{figure}[ht]
\begin{center}
\epsfig{file=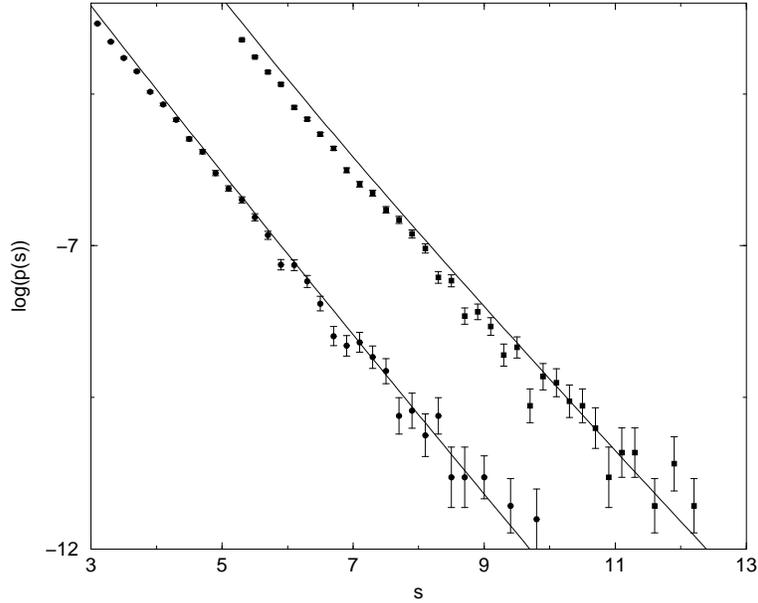, angle=-90,   width=10  cm}
\end{center}
\caption{The nearest-neighbor distribution for a billiard with Dirichlet 
  boundary  conditions. Squares and circles correspond respectively to 
$\lambda'=1$ and $\lambda'=10$. Solid lines: theoretical asymptotics
(\ref{psfini}). Upper curves shifted to the right by 2 units.}
\label{psndir1}
\end{figure}
To better understand the asymptotics of the nearest-neighbor distribution
we present in Fig.~\ref{pentes} the functions which determine the
exponential decrease of $P(s)$. Different curves in this figure correspond
to the following functions: $-\ei(x)$ (thick solid line), $\exp x$ (thin
solid line), $-<r\ei (rx)>$ (thick dashed line), and $<\exp rx>$ (thin
dashed line). 
\begin{figure}[ht]
\begin{center}
\epsfig{file=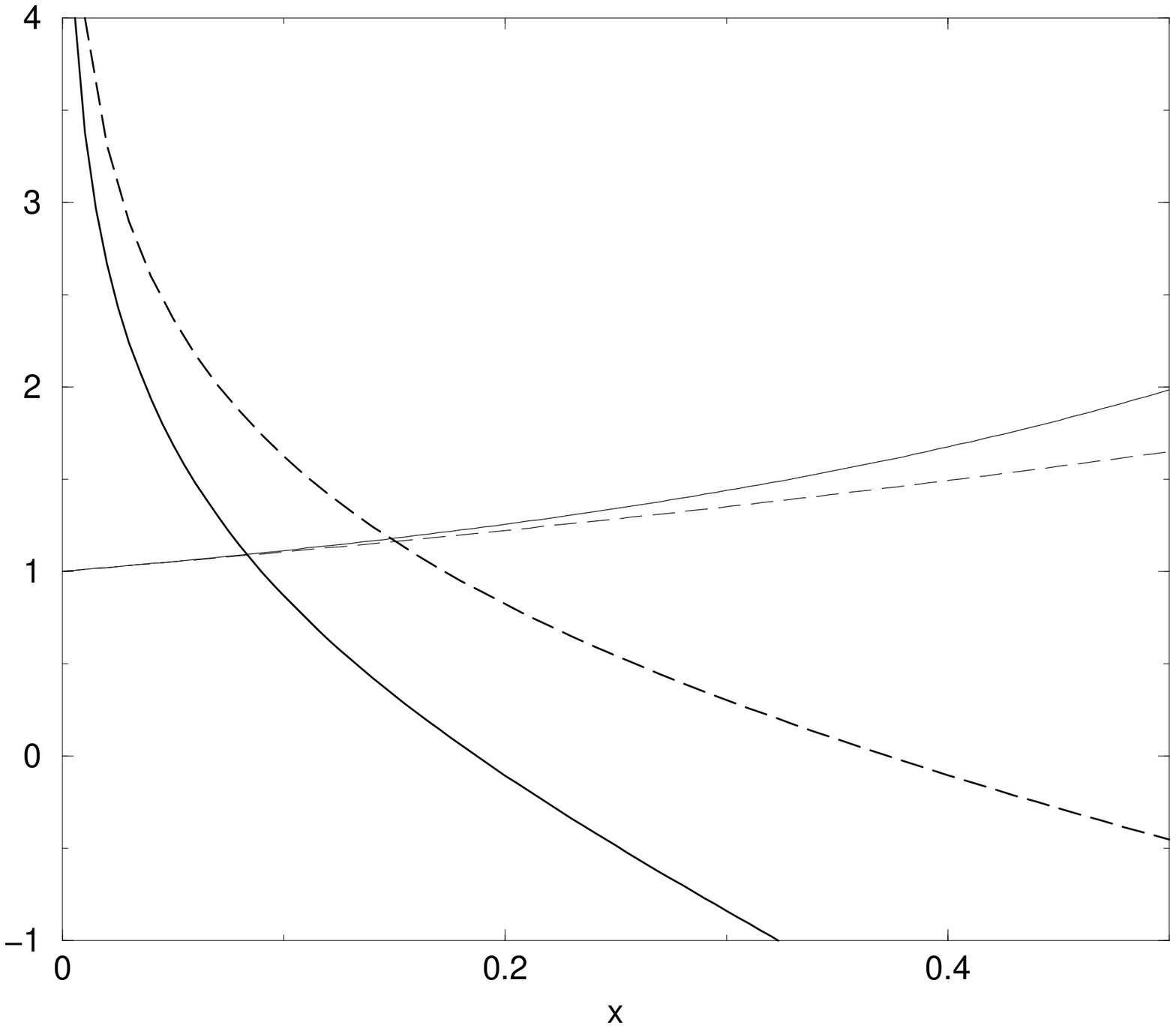,width=10cm, angle=-90}
\end{center}
\caption{Functions $-\ei(x)$ (thick solid line), $-<r\ei(r x)>$ (thick dashed
  line),  $e^{x}$ (thin solid line), and $<e^{r x}>$ (thin dashed line).}
\label{pentes}
\end{figure}

For periodic boundary conditions
the intersection of the horizontal line having the ordinate $1/\lambda'$
with the graph of $-\ei(x)$  gives the value of $\vsp$.
The point of intersection of the vertical line going through $v_{sp}$ with
the graph of $e^{x}$ determines the exponent in Eq.~(\ref{psfini}). 
For Dirichlet boundary conditions one should use the same procedure but with
the functions $-<r\ei(r x)>$ and $<e^{r x}>$.

\section{The $n$-th nearest-neighbor spacing distribution}\label{nnearest}

In the previous Sections we compute the nearest-neighbor distribution, $P(s)$, for a
rectangular billiard with a small-size scattering center inside, i.e. the
probability that 2 levels are separated by a distance $s$ with no
levels in between. In this Section we generalize the formalism to compute the
$n$-th nearest-neighbor distribution, $P_n(s)$, which is defined as
the probability that 2 levels at distance $s$ are separated by $n$ levels,
for $n\geq 1$. 

Our starting point is the expression (\ref{pomega}) which is valid for any
sequence $\sigma_k$ of 0 and 1. To obtain the $n$-th nearest-neighbor 
distribution one
should sum over all sequences of length $N$ with exactly $n+1$ zeros.
Performing the same steps as in Section~\ref{nnd} and taking into account
that when $N\rightarrow \infty$ we have $C_{N}^n\rightarrow N^n/n!$ one obtains
\begin{eqnarray}
&&P_{n}(s)=e^{-s}\int_{-\infty}^{\infty}\frac{d\alpha_1 d\alpha_2}{(2\pi)^2}
[\frac{s^{n+1}}{(n+1)!}(1-<\tilde{J}_0(\alpha r)>)^{n+1} 
V_{\phi_1,\phi_1}(\alpha)
\nonumber\\
&&+\frac{s^{n}}{n!}(1-<\tilde{J}_0(\alpha r)>)^{n} 
(V_{\phi_0,\phi_1}(\alpha)+V_{\phi_1,\phi_0}(\alpha))
\nonumber\\
&&+\frac{s^{n-1}}{(n-1)!}(1-<\tilde{J}_0(\alpha r)>)^{n-1}
 V_{\phi_0,\phi_0}(\alpha)] 
 e^{-s(<\tilde{J}_1(\alpha r)>+i(\alpha_1+\alpha_2)/\lambda')},
\label{psnfinal}
\end{eqnarray}
with all functions defined as above. 

As a consistency check one can verify  that the sum over all $n$ 
coincides with the exact expression of the 2-point correlation function
derived in \cite{singular}  
\be 
R_2(s)=\sum_{n=0}^{\infty}P_n(s).
\end{equation}

Similarly to Section~\ref{analytical}  the different terms in Eq.~(\ref{psnfinal})
can be classified according to their analytical properties. Three groups
of terms appear. First, terms which have the same analytical
properties as the function $\tilde{J}_1(\alpha)$ in the exponent. Exactly as
it was done in Section~\ref{analytical} one can argue that their
contribution is  zero
because one can freely move the integration contour to infinity. The second
group consists of terms which are singular on one variable but have `good'
analytical properties on the other variable of integration. The large $s$
asymptotics of such terms can be calculated as in Section~\ref{asymbeh} by
computing the leading terms over the former variable and shifting the
contour of integration over the latter variable into the complex
plane till it reaches the saddle point. The asymptotics of these terms
will be proportional to $\exp [ -se^{\vsp}]$ as for the nearest-neighbor
distribution. Finally, the third group  (which does  not exist for
$P(s)$) includes terms which are singular (i.e. have analytical properties
different from that of $\tilde{J}_1(\alpha)$) for both variables of
integration. In this case no deformation of the integration contour is possible,
the region of small $\alpha_1$ and $\alpha_2$ will be important and the
asymptotic result will be proportional to $\exp (-s)$. As $\vsp>0$ it is the
last group of terms which will dominate the asymptotics of $P_n(s)$ when
$s\rightarrow \infty$. 

From formula  (\ref{smallphi})  of Appendix
it follows that, when both variables $\alpha_1$ and $\alpha_2$ are small,
the functions $V_{\phi_i, \phi_j}$ will be equivalent to their singular parts
\begin{eqnarray}
&&V_{\phi_1,\phi_1}(\alpha)\simeq \frac{1}{\alpha_1\alpha_2},\;\;
V_{\phi_1,\phi_0}(\alpha)\simeq
-\frac{2\pi i}{\alpha_1}\delta (\alpha_2),\nonumber\\
&&V_{\phi_1,\phi_0}(\alpha) \simeq
\frac{2 \pi i}{\alpha_2}\delta (\alpha_1),\;\;
V_{\phi_0,\phi_0}(\alpha)\simeq (2\pi)^2\delta(\alpha_1)\delta(\alpha_2).
\end{eqnarray}
The singular terms in $(1-<\tilde{J}_0(r\alpha)>)^n$ are given by the small
$\alpha$ behavior in Eq.~(\ref{smallJ0}). Keeping only the terms with a
singularity in 
$[1+<\tilde{J}_1(r\alpha)>-(<\tilde{J}_0(r\alpha)>+<\tilde{J}_1(r\alpha)>)]^n$
we get
\be
(1-<\tilde{J}_0(r\alpha)>)^n=1
-n\pi(|\alpha_1|+|\alpha_2|)
+n(n-1)\pi^2|\alpha_1| |\alpha_2|.
\end{equation}
(We recall that the terms coming from $\tilde{J}_1$ vanish.)
The first term in this expression is the dominant regular contribution, the 
second one is the
dominant contribution singular in one variable, and the third term is the
dominant contribution singular in both variables.

Combining all the above expressions together and using (\ref{smallJ1}) we obtain
\begin{eqnarray}
&&P_{n}(s)=\frac{s^{n-1}}{(n-1)!}e^{-s}\int_{-\infty}^{\infty}
\frac{d\alpha_1 d\alpha_2}{(2\pi)^2}
[\pi^2s^2\mbox{sgn}(\alpha_1)\mbox{sgn}(\alpha_2)
\nonumber\\
&&+2\pi^2 i s\  \mbox{sgn}(\alpha_1)\delta (\alpha_2) 
-2\pi^2 i s\  \mbox{sgn}(\alpha_2)\delta (\alpha_1)
+(2\pi)^2\delta(\alpha_1)\delta(\alpha_2)]\\
&&\times \exp -s[\frac{\pi}{2}(|\alpha_1|+|\alpha_2|)
-i\alpha_{1}(\log |\alpha_{1}|+g_{-})
+i\alpha_{2} (\log |\alpha_{2}|+g_{+})],
\nonumber
\end{eqnarray}
where
\be
g_{\pm}=<r\log r>+\gamma -1 \pm \frac{1}{\lambda'}.
\label{gpm}
\end{equation}
After simple calculations we obtain the large $s$ asymptotics of the $n$-th
nearest-neighbor distribution
\be 
P_n(s)=\frac{s^{n-1}}{(n-1)!}e^{-s}[1-f(\log s -g_{+})][1-f(\log s - g_{-})],
\label{asymppns}
\end{equation}
where the function $f(y)$ is defined by  Eq.~(\ref{f}) (see also
Fig.~\ref{function}).

The only difference between periodic and Dirichlet boundary conditions 
is in the constant $g_{\pm}$ (\ref{gpm}) where the term
$<r\log r>=2(1-\log 2)$ is added for the latter. 

As above when $s\rightarrow \infty$ the function $f(\log s -g_{\pm})$ goes
to zero and the true asymptotics of the $n$-th nearest-neighbor distribution
with $n\geq 1$ for all boundary conditions is 
\be
P_n(s)=\frac{s^{n-1}}{(n-1)!}e^{-s},
\label{trueasymp}
\end{equation}
i.e. it coincides with the $(n-1)$-th nearest-neighbor distribution for the
Poisson distribution. A simple physical explanation of this result is the
following. 

We are interested in the solutions of Eq.~(\ref{equN}) when unperturbed
levels, $e_j$, are independent random variables. Among all configurations
of  $e_j$ there are cases where two  unperturbed levels $e_1$ and $e_2$ are very
close to each other. In such a case Eq.~(\ref{equN}) reduces to 2 terms
\be
\frac{r_1}{E-e_1}+\frac{r_2}{E-e_2}=0,
\end{equation}
which has a simple solution
\be
E=\frac{e_1r_2+e_2r_1}{r_1+r_2}.
\label{86}
\end{equation}
As we assume that the difference $e_1-e_2$ is very small, the value of $E$
will also be very close to both unperturbed levels, $e_1$ and $e_2$. 
The equation (\ref{86}) can be reversed
and having one unperturbed level, say $e_1$, and  new level, $E$, very close
to it, one can always find the position of another unperturbed level, $e_2$, to
fulfill Eq.~(\ref{equN}) in that approximation. We shall call 2 very close 
unperturbed levels with a
new level inside a dipole  configuration. Now let us consider the
probability that 2 unperturbed levels  $e_1$ and $e_2$ are at the distance $s$ 
with $n-1$ unperturbed levels inside this interval. As unperturbed levels are
independent this probability is given by Eq.~(\ref{trueasymp}). 

We represent in Fig.~\ref{schema} a possible configuration for 2
unperturbed levels (short thin vertical lines) separated by a large distance
$s=e_1-e_2$ with $n-1$ levels between them.  For any such configuration there exists a configuration with
2 new energy levels at $E_2$ and $E_1$ very close to $e_2$ and $e_1$
respectively. This is true because, as we have pointed out, it is possible to
construct two unperturbed energy levels $e_2'$ and $e_1'$ (indicated by
dashed lines in Fig.~\ref{schema}) such that two pairs
$(e_2',e_2)$, $(e_1',e_1)$ form the dipoles. For that configuration it is
clear that $E_2$ and $E_1$ are two perturbed levels with $n$
perturbed levels in between. When $s \rightarrow \infty$ it is physically clear
that the other levels will not influence these dipole configurations which
explains  Eq.~(\ref{trueasymp}).
\begin{figure}[ht]
\begin{center}
\epsfig{file=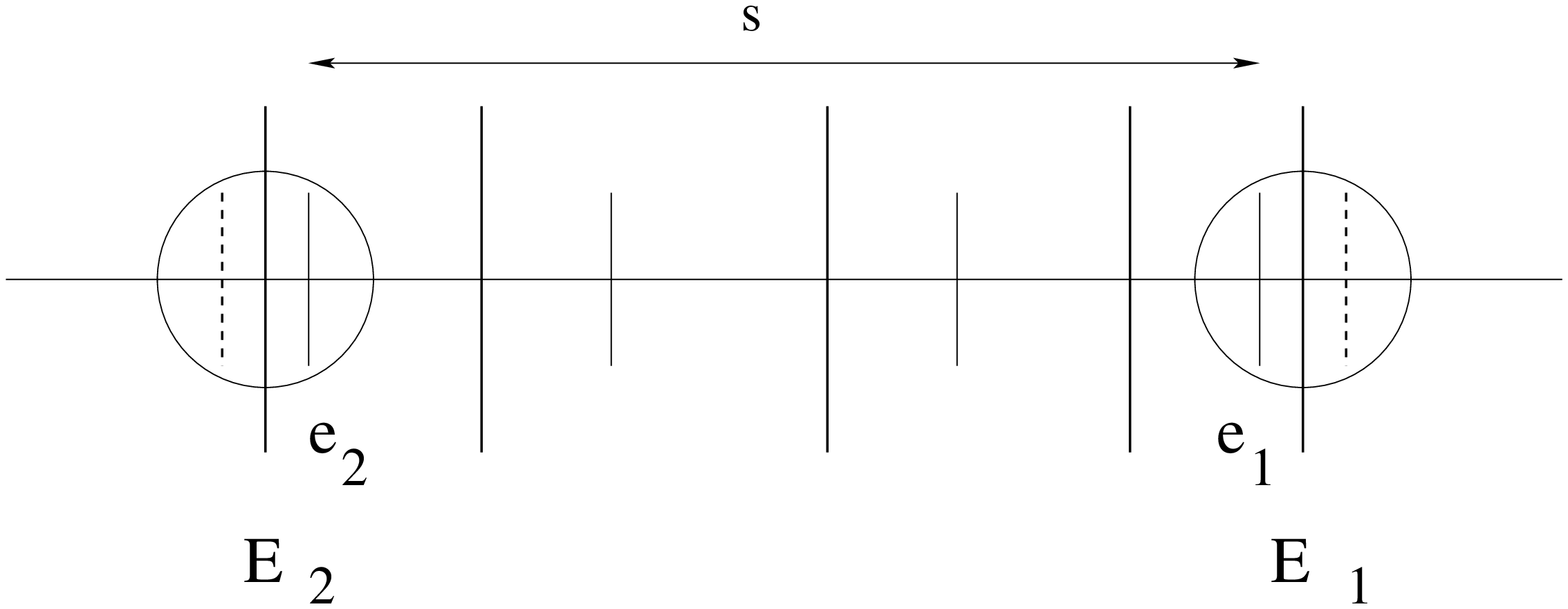, width=10cm, angle=-90}
\end{center}
\caption{Schematic representation of the dominant contribution 
  to the $n$-th nearest-neighbor distribution. Thin vertical lines are
   unperturbed energy levels. Thick lines represent new energy levels. 
   2 vertical dashed lines indicate 2  unperturbed energy levels which are
   added  to construct encircled dipole configurations.}
\label{schema}
\end{figure}

It is clear that this reasoning cannot be applied to the nearest-neighbor
distribution (as it requires at least 2 unperturbed levels inside the
interval $s$) and the asymptotics of $P(s)$ given by Eq.~(\ref{trueasympps})
is quite different from Eq.~(\ref{trueasymp}).

To check the asymptotic formula (\ref{trueasymp}) we compute
numerically the $n$-th nearest-neighbor distributions till $n=9$ for  a
rectangular billiard with size $4\times \pi$  with periodic boundary
conditions and for different coupling constants. First, we find the best fit
of the integrated $n$-th nearest-neighbor distribution
$N_{n}(s)=\int_0^sP_n(t)dt$ in the
form $a \exp(b s) s^{c}$. In Fig.~\ref{parametresper}
we plot values of $b$ and $c$ obtained by this fit. The lower curve (dots)
shows the values of the constant $b$, which as expected is the same for all
values of $n$ and is equal with a good precision to $-1$. The upper curve
(squares) shows the exponent $c$ which according to $(\ref{asymppns})$ is
expected to be $n-1$. This is indeed the case for the higher values of $n$.
The small deviations from this expected value for the lowest $n$ comes from the
fact that the function $f$ has to be taken into account since for large $s$,
$f(\log s-g_{\pm})$ behaves like $1/\log s$.
\begin{figure}[ht]
\begin{center}
\epsfig{file=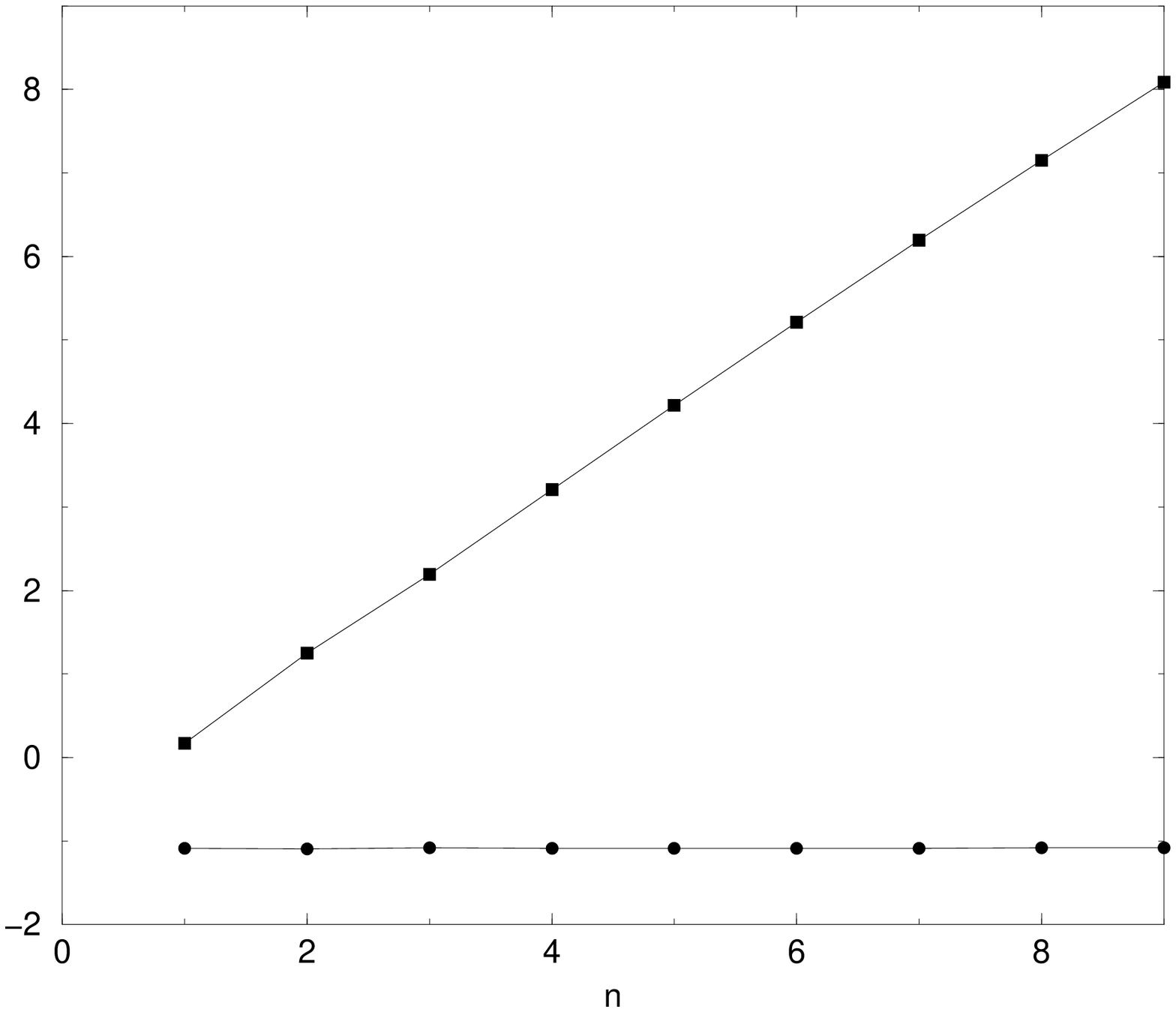 ,width=10cm,angle=-90}
\end{center}
\caption{Values of $b$ (dots) and $c$ (squares) in a fit of $N_{n}(s)$ for
  periodic boundary conditions under the form $a \exp(b s) s^{c}$.}
\label{parametresper}
\end{figure}
To illustrate accuracy of Eq.~(\ref{asymppns}) we present in
Fig.~\ref{p2_8per} the results of numerical computations for $P_2(s)$ and
$P_8(s)$ for a billiard with periodic boundary conditions and renormalized
coupling constant, $\lambda'=1$. Error bars in this figure indicate
statistical errors.   
\begin{figure}[ht]
\begin{center}
\epsfig{file=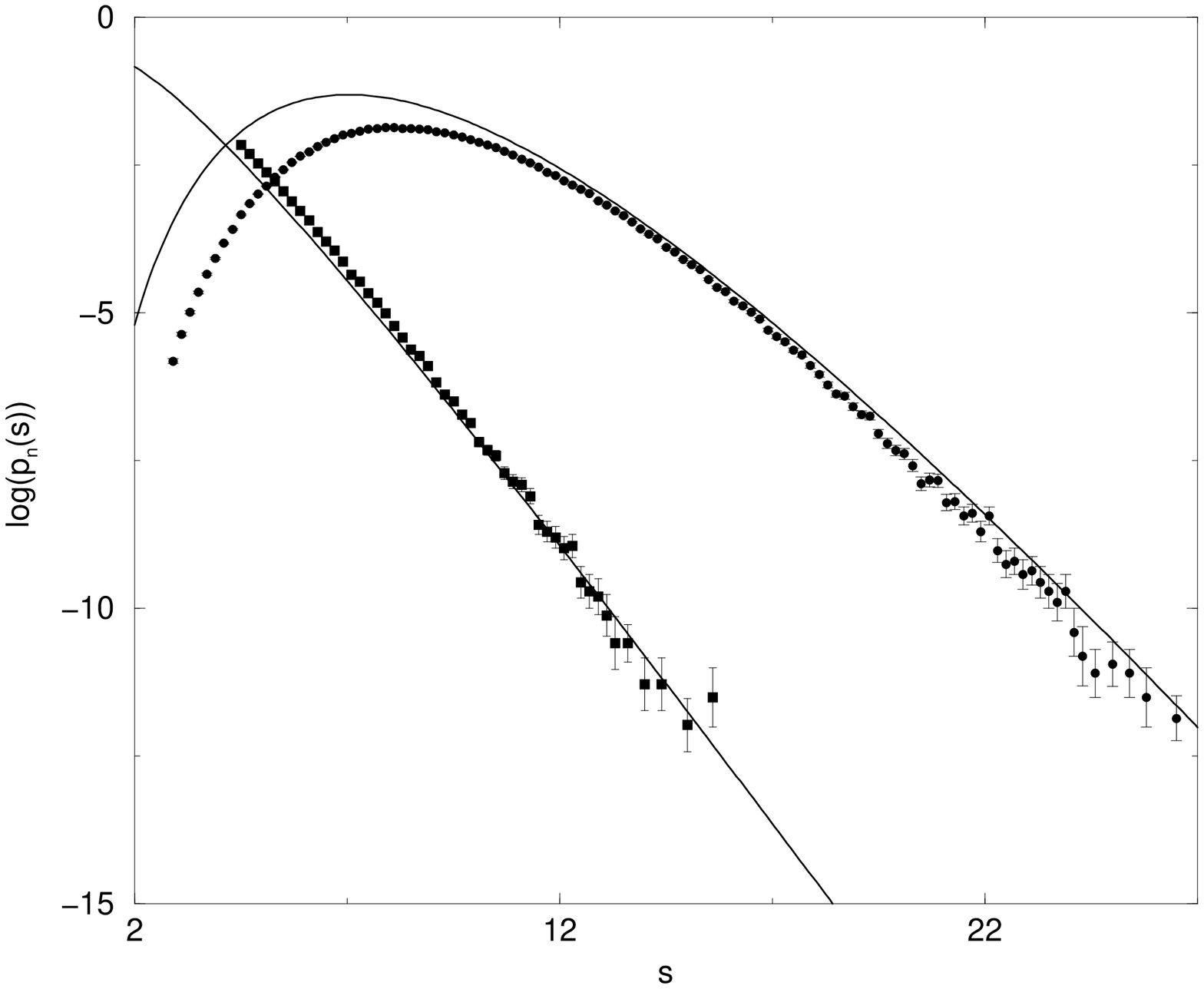 ,width=10cm,angle=-90}
\end{center}
\caption{$P_{2}(s)$ (the left curve) and $P_{8}(s)$ (the right curve) 
  for a billiard with periodic boundary conditions.}
\label{p2_8per}
\end{figure}
It is clear that the asymptotic formula (\ref{asymppns}) describes well
large $s$ behavior of $P_n(s)$.  

The same checks have been performed for billiards with Dirichlet boundary
conditions. The results are presented in Fig.~\ref{parametresdir} and
Fig~\ref{p2_8dirichlet}. 
\begin{figure}[ht]
\begin{center}
\epsfig{file=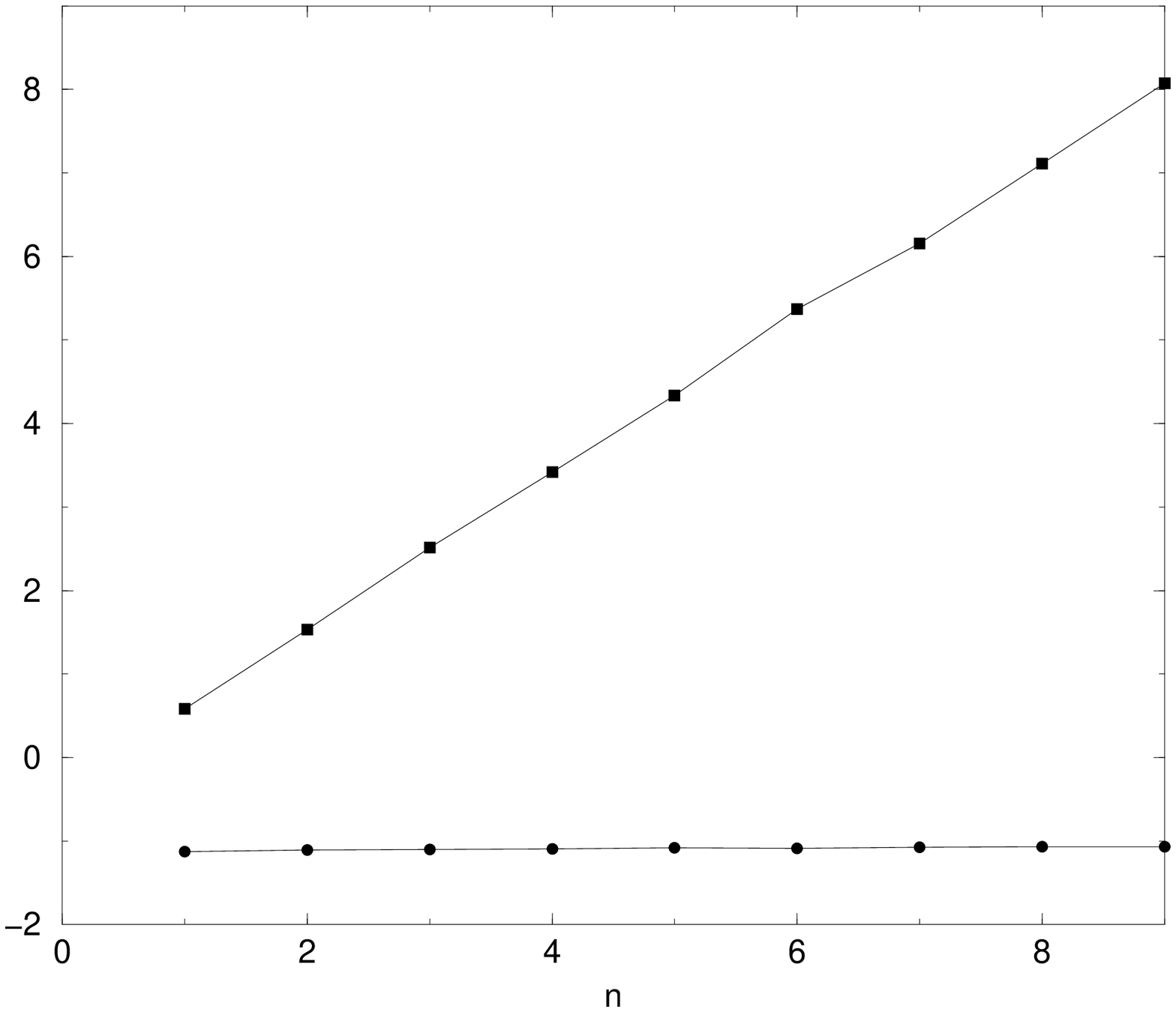 ,width=10 cm,angle=-90}
\end{center}
\caption{Values of $b$ (dots) and $c$ (squares) in a fit of $N_{n}(s)$ for
  Dirichlet boundary conditions under the form $a \exp(b s) s^{c}$.}
\label{parametresdir}
\end{figure}
\begin{figure}[ht]
\begin{center}
\epsfig{file=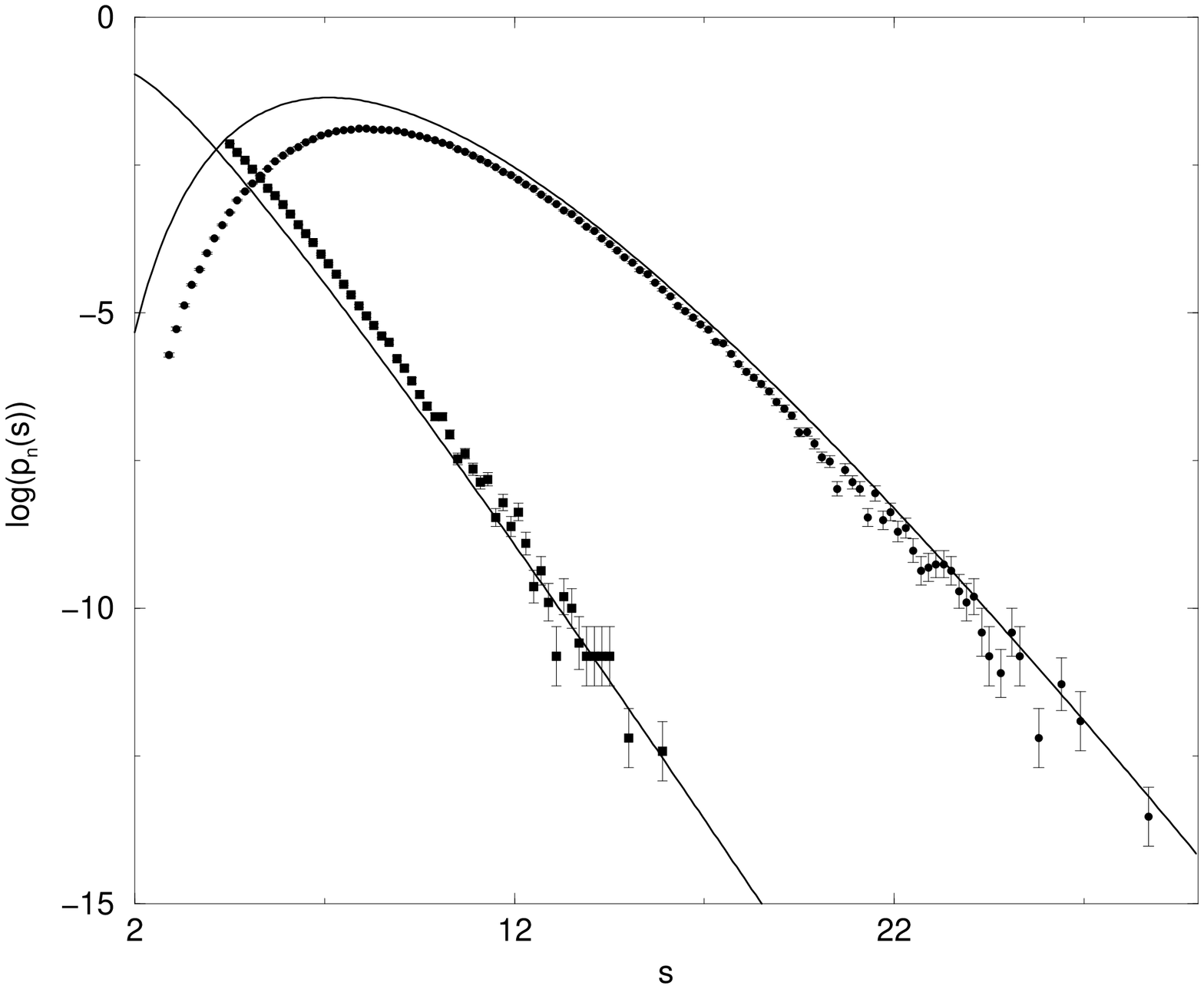 ,width=10 cm,angle=-90}
\end{center}
\caption{$P_{2}(s)$ (the left curve) and $P_{8}(s)$ (the right curve) 
  for a billiard with Dirichlet boundary condition}
\label{p2_8dirichlet}
\end{figure}
Again the theoretical asymptotics (\ref{asymppns}) is  in a good agreement
with numerical results.

\section{Conclusion}

The starting point of our investigation is Eq.~(\ref{equN}) 
\be
\lambda \sum_{j=1}^{N} \frac{r_j}{E-e_j}=1.
\end{equation}
We assume that (i) all $e_j$ are independent random variables, (ii) the
residues $r_j$ are real positive,  and we compute the $n$-th nearest-neighbor distribution
$P_n(s)$ of the solutions, $E$, in the limit $N\rightarrow \infty$.
The exact formulas are quite cumbersome and we dwell on the asymptotic
behavior of $P_n(s)$ at large $s$. Our main results are the following.

The asymptotics of the nearest-neighbor distribution, $P(s)$, is given by
Eq.~(\ref{trueasympps}) and has the form 
\be
P(s)=\frac{(<e^{rv_{sp}}>)^2}{\sqrt{2\pi<re^{rv_{sp}}>v_{sp}}}
    \frac{e^{-s<e^{rv_{sp}}>}}{\sqrt{s}}.
\label{conclps}
\end{equation}
Here $v_{sp}$ is determined by the equation
\be
-<r\ei (rv_{sp})> =\frac{1}{|\lambda'|},
\end{equation}
where $\ei(x)$ is the usual exponential integral (\ref{EI}) and $\lambda'$
is the renormalized coupling constant (\ref{lprime}). The notation $<f(r)>$
indicates the mean value over all residues. For periodic boundary conditions
$<f(r)>=f(1)$. For Dirichlet conditions $<f(r)>$ depends on the ratios of
the coodinates of the scatterer to the corresponding sides. When these
ratios are non-commensurable irrational numbers $<f(r)>$ is defined in
(\ref{fmean}). When they are rational numbers $<f(r)>$ should be computed as
in (\ref{fmeanrational}).

The $n$-th nearest-neighbor distribution with $n\geq 1$ when 
$s\rightarrow \infty$ has the  following asymptotics
\be
P_n(s)=\frac{s^{n-1}}{(n-1)!}e^{-s}
\end{equation}
which  depends neither on the residues nor on the boundary conditions.
For finite values of $s$ there are slowly decreasing corrections to these
formulas indicated in Eqs.~(\ref{asympps}) and (\ref{asymppns}) which are 
important for accurate comparison with results of numerical calculations. 

The above results together with the results of Ref.~\cite{singular} prove
that spectral statistics of generic rectangular billiards with a small-size
scattering center inside is of special (intermediate) type characterized by
2 important properties: (i) level repulsion at small $s$  and (ii)
exponential decrease of the nearest-neighbor distribution at large $s$. For
the authors' knowledge this is the first example of a dynamical system where 
the intermediate character of the spectral statistics can be proved
rigorously. 

\section*{Appendix}

The purpose of this Appendix is the computation of two  main integrals 
\be
\tilde{J}_1(\alpha_1,\alpha_2)=
 (\int\limits_{-\infty}^{0}+\int\limits_{1}^{\infty} )
[1-\exp (i\frac{\alpha_1}{1-e}-i\frac{\alpha_2}{e})]  d e,
\label{aj1}
\end{equation}
and 
\be
\tilde{J}_0(\alpha_1,\alpha_2)=\int_{0}^{1}
[1-\exp (i\frac{\alpha_1}{1-e}-i\frac{\alpha_2}{e})]  d e.
\label{aj0}
\end{equation}
As in \cite{singular} we first  find the difference of the derivatives over
$\alpha_1$ and $\alpha_2$  
\be
(\da-\db)\tilde{J}_1(\alpha_1,\alpha_2)=
-i(\int\limits_{-\infty}^{0}+\int\limits_{1}^{\infty})
\frac{de}{e(1-e)}\exp(i\frac{\alpha_1}{1-e}-i\frac{\alpha_2}{e}),
\end{equation}
where $\partial_i$ denotes the partial derivative with
respect to the $\alpha_i$. Changing the variable 
$e=t/(1+t)$ one gets
\be
(\da-\db)\tilde{J}_1(\alpha_1,\alpha_2)=
e^{i(\alpha_1-\alpha_2)}\phi_1(\alpha_1,\alpha_2),
\label{difference}
\end{equation}
where
\be
\phi_1(\alpha_1,\alpha_2)=i\int_{0}^{\infty}\frac{dt}{t}
\exp(-i\alpha_1 t+i\frac{\alpha_2}{t}).
\end{equation}
The last integral is well defined  for complex $\alpha$ when 
$\im(\alpha_1)<0$ and $\im(\alpha_2)>0$. In this domain (see \cite{bateman})
\be
\phi_1(\alpha_1,\alpha_2)=
2 i K_0(2\sqrt{\alpha_1 \alpha_2})
\label{phi1}
\end{equation}
where $K_0(x)$ is the modified Bessel function of the third kind.
At real $\alpha$ the limiting function is discontinuous and depends on the
sign of $\alpha$
\be
\phi_1(\alpha_1,\alpha_2)=\left \{ \begin{array}{rl}
-\pi H^{(1)}_0(2\sqrt{-\alpha_1\alpha_2})
  &\mbox{when }\;\alpha_1<0,\;\alpha_2>0\\
2iK_0(2\sqrt{\alpha_1\alpha_2})
  &\mbox{when }\;\alpha_1>0,\;\alpha_2>0\\
\pi H^{(2)}_0(2\sqrt{-\alpha_1\alpha_2})
  &\mbox{when }\;\alpha_1>0,\;\alpha_2<0\\
  2i K_0(2\sqrt{\alpha_1\alpha_2})
  &\mbox{when }\;\alpha_1<0,\;\alpha_2<0
\end{array}\right . .
\label{realphi1}
\end{equation}
The function $\tilde{J}_0(\alpha)$ obeys the similar equation
\be
(\da-\db)\tilde{J}_0(\alpha_1,\alpha_2)=
e^{i(\alpha_1-\alpha_2)}\phi_0(\alpha_1,\alpha_2),
\end{equation}
where
\be
\phi_0(\alpha_1,\alpha_2)=-i\int_{0}^{\infty}\frac{dt}{t}
\exp (i\alpha_1 t-i\frac{\alpha_2}{t})=
-2iK_0(2\sqrt{\alpha_1 \alpha_2}),
\label{phi0}
\end{equation}
which is defined in the region $\im(\alpha_1)>0$ and $\im(\alpha_2)<0$. At
real $\alpha$, $\phi_0(\alpha)=\phi_1^*(\alpha)$ or explicitly
\be
\phi_0(\alpha_1,\alpha_2)=\left \{\begin{array}{rl}
-\pi H^{(2)}_0(2\sqrt{-\alpha_1\alpha_2})
  &\mbox{when }\;\alpha_1<0,\;\alpha_2>0\\
-2iK_0(2\sqrt{\alpha_1\alpha_2})
  &\mbox{when }\;\alpha_1>0,\;\alpha_2>0\\
\pi H^{(1)}_0(2\sqrt{-\alpha_1\alpha_2})
  &\mbox{when }\;\alpha_1>0,\;\alpha_2<0\\
  -2iK_0(2\sqrt{\alpha_1\alpha_2})
  &\mbox{when }\;\alpha_1<0,\;\alpha_2<0
\end{array}\right . .
\label{realphi0}
\end{equation}
We note also the expression for the sum of $\phi_{\sigma}(\alpha)$
\be
\phi_{1}(\alpha_1,\alpha_2)+\phi_{0}(\alpha_1,\alpha_2)=
\pi(\mbox{sgn}(\alpha_1)-\mbox{sgn}(\alpha_2))J_0(2\sqrt{-\alpha_1\alpha_2}),
\label{sumphi}
\end{equation}
where sgn$(x)$ denotes the sign of $x$.

The knowledge of $\phi_{\sigma}(\alpha)$ permits to write down a
linear partial derivative equation for $J_{\sigma}(\alpha)$
\be
(\da-\db)\tilde{J}_{\sigma}(\alpha_1,\alpha_2)=Z_{\sigma}(\alpha_1,\alpha_2)
\end{equation}
where
\be
Z_{\sigma}(\alpha_1,\alpha_2)=
e^{i(\alpha_1-\alpha_2)}\phi_{\sigma}(\alpha_1,\alpha_2).
\end{equation}
The  general solution of this equation has the form
\be
\tilde{J}_{\sigma}(\alpha_1,\alpha_2)=
\tilde{J}_{\sigma}(0,\alpha_1+\alpha_2)+\int_{0}^{\alpha_1}
Z_{\sigma}(t,\alpha_1+\alpha_2-t) d t.
\end{equation}
The initial values $\tilde{J}_{\sigma}(0, \alpha)$ can be computed directly from 
the definitions (\ref{aj1}) and (\ref{aj0}). By changing the variable $e$ to 
$t=-1/e$ one gets
\be
\tilde{J}_1(0,\alpha)=
(\int_{-\infty}^{0}+\int_{1}^{\infty})
( 1-e^{-i\alpha/e} )
  d e=\vpint_{-1}^{\infty}\frac{d t}{t^2}
(1-e^{i\alpha t}),
\end{equation}
where ${\mathbf -}\hspace{-1em}\int$ is the principal value of the integral. Similarly
\be
\tilde{J}_0(0,\alpha)=
\int_{0}^{1}( 1-e^{-i\alpha/e})d e=\int_{1}^{\infty}\frac{d t}{t^2}
(1-e^{-i\alpha t}).
\end{equation}
The final expressions for $\tilde{J}_{\sigma}(\alpha)$ are the following
\be
\label{fonctionJ1}
\tilde{J}_1(\alpha_1,\alpha_2)=\vpint_{-1}^{\infty}\frac{d t}{t^2}
(1-e^{i(\alpha_1+\alpha_2) t})
+\int_{0}^{\alpha_1}e^{i(2 t-\alpha_1-\alpha_2)}
\phi_1(t,\alpha_1+\alpha_2-t) d t,
\end{equation}
and 
\be
\label{fonctionJ0}
\tilde{J}_0(\alpha_1,\alpha_2)=\int_{1}^{\infty}\frac{d t}{t^2}
(1-e^{-i(\alpha_1+\alpha_2) t})
+\int_{0}^{\alpha_1}e^{i(2 t-\alpha_1-\alpha_2)}
\phi_0(t,\alpha_1+\alpha_2-t) d t.
\end{equation}
According to Eqs.~(\ref{aj0}) and (\ref{aj1}) the functions
$\tilde{J}_0(\alpha)$ and $\tilde{J}_1(\alpha)$ considered as
functions of complex $\alpha_1$ and $\alpha_2$ are analytical
functions in different regions: $\im(\alpha_1)>0$ and $\im(\alpha_2)<0$
for the former and  $\im(\alpha_1)<0$ and  $\im(\alpha_2)>0$ for the latter.
They can be continued to complex planes with the cuts represented 
in Fig.~\ref{cut}a and \ref{cut}b for $\tilde{J}_1$ and
in Fig.~\ref{cut}c and \ref{cut}d for $\tilde{J}_0$.
At real values of $\alpha_1$ and  $\alpha_2$ they have a 
discontinuity along the axis $\alpha_1=0$ and $\alpha_2=0$.  

The sum  $\tilde{J}_0(\alpha)+\tilde{J}_1(\alpha)$ obeys an equation similar to
(\ref{difference}) where instead of function $\phi_1(\alpha)$ one
substitutes the sum (\ref{sumphi})
\be
(\da-\db)[\tilde{J}_1(\alpha_1,\alpha_2)+\tilde{J}_0(\alpha_1,\alpha_2)]=
e^{i(\alpha_1-\alpha_2)}Z(\alpha_1,\alpha_2)
\end{equation}
where
\be
Z(\alpha_1,\alpha_2)=\pi(\mbox{sgn}(\alpha_1)-\mbox{sgn}(\alpha_2))
J_0(2\sqrt{-\alpha_1\alpha_2}).
\end{equation}
The solution of these equations can be done exactly as above and it can be 
represented as a sum of two discontinuous functions
\be
\tilde{J}_0(\alpha_1,\alpha_2)+\tilde{J}_1(\alpha_1,\alpha_2)=
\mbox{sgn}(\alpha_1)R(\alpha_1,\alpha_2)+
\mbox{sgn}(\alpha_2)R^{*}(\alpha_2,\alpha_1),
\label{sumj}
\end{equation}
where
\be
R(\alpha_1,\alpha_2)=\pi
\int_{0}^{\alpha_1}J_{0}(2\sqrt{-t(\alpha_1+\alpha_2-t)})e^{i(2t-\alpha_1-\alpha_2)}dt.
\label{R}
\end{equation}
One can check that this expression coincides with Eq.~(A24) of \cite{singular}.

The following useful symmetry properties can be checked directly from the 
definitions (\ref{aj1})-(\ref{aj0})
\begin{eqnarray}
&&\tilde{J}_{\sigma}(\alpha_2,\alpha_1)=\tilde{J}_{\sigma}(-\alpha_1,-\alpha_2)=
\tilde{J}_{\sigma}^{*}(\alpha_1,\alpha_2),\nonumber\\
&&\phi_{\sigma}(\alpha_2,\alpha_1)=\phi_{\sigma}(-\alpha_1,-\alpha_2)=
-\phi_{\sigma}^{*}(\alpha_1,\alpha_2),\label{symmetry}\\
&&R(-\alpha_1,-\alpha_2)=-R^{*}(\alpha_1,\alpha_2).
\nonumber
\end{eqnarray}

For further references we present the  behavior of
$\tilde{J}_{\sigma}(\alpha)$ and $\phi_{\sigma}(\alpha)$ at small real $\alpha$
\begin{eqnarray}
&&\tilde{J}_1(\alpha_1,\alpha_2)=\frac{\pi}{2}(|\alpha_1|+|\alpha_2|)
\label{smallJ1}\\
&&+i\alpha_{1}(1-\gamma-\log |\alpha_{1}|)
-i\alpha_{2} (1-\gamma-\log |\alpha_{2}|),
\nonumber\\
&&\tilde{J}_0(\alpha_1,\alpha_2)=\frac{\pi}{2}(|\alpha_1|+|\alpha_2|)
\label{smallJ0}\\
&&-i\alpha_{1}(1-\gamma-\log |\alpha_{1}|)
+i\alpha_{2} (1-\gamma-\log |\alpha_{2}|),
\nonumber
\end{eqnarray}
and
\be
\phi_1(\alpha_1,\alpha_2)=
\frac{\pi}{2}(\mbox{sgn}(\alpha_1)-\mbox{sgn}(\alpha_2))
-i(2\gamma + \log |\alpha_1|+\log |\alpha_2|),
\label{smallphi}
\end{equation}
\be
\phi_0(\alpha_1,\alpha_2)=
\frac{\pi}{2}(\mbox{sgn}(\alpha_1)-\mbox{sgn}(\alpha_2))
+i(2\gamma + \log |\alpha_1|+\log |\alpha_2|),
\end{equation}
where $\gamma$ is the Euler constant.

Due to the above mentioned analytical properties these expressions \\
(though they have apparent discontinuities) can be
rewritten as  following analytical functions
\be
\tilde{J}_1(\alpha_1,\alpha_2)=
i\alpha_{1}(1-\gamma-\log(i\alpha_{1}))
-i\alpha_{2} (1-\gamma-\log (-i\alpha_{2})),
\end{equation}  
\be
\tilde{J}_0(\alpha_1,\alpha_2)=
-i\alpha_{1}(1-\gamma-\log(-i\alpha_{1}))
+i\alpha_{2} (1-\gamma-\log (i\alpha_{2})),
\end{equation}
and
\be
\phi_1(\alpha_1,\alpha_2)=
-i(2\gamma + \log (i\alpha_1)+ \log(-i\alpha_2)),
\label{phismall}
\end{equation}
\be
\phi_0(\alpha_1,\alpha_2)=
i(2\gamma + \log (-i\alpha_1)+ \log(i\alpha_2)).
\end{equation}
where we assume the usual definition of the logarithmic function with a cut
along the real negative axis. 

Similarly as it was done in \cite{singular} for the sum
$\tilde{J}_1(\alpha)+\tilde{J}_0(\alpha)$ one can also obtain higher order terms of
the expansions of $\tilde{J}_{\sigma}(\alpha)$ in power of $\alpha$.
Both functions $\tilde{J}_1$ and $\tilde{J}_0$ given by Eqs.~(\ref{fonctionJ1}), 
(\ref{fonctionJ0}) contain terms proportional to $\log (\alpha_1+\alpha_2)$
but one can check by direct series expansions  that in the corresponding sums 
these terms  all cancel and the only logarithmic contributions to these
functions are the ones  presented in Eqs.~(\ref{smallJ1}) and
(\ref{smallJ0}) as it should be to ensure the  analytical properties of
$\tilde{J}_{\sigma}(\alpha)$.

\end{document}